\documentclass[12pt,letterpaper]{article}
\usepackage{epsfig,rotating,setspace,latexsym,amsmath,epsf,amssymb,amsfonts,bm,theorem,cite,caption,subcaption,enumerate,longtable,accents,algorithm,graphicx,epsf,authblk,epstopdf,url,color,multirow,dsfont,mathtools,comment,optidef}
\usepackage[noend]{algpseudocode}

\DeclareMathOperator*{\argmin}{arg\,min}

\newtheorem{theorem}{Theorem}

\newtheorem{definition}{Definition}
\newtheorem{remark}{Remark}

\newtheorem{example}{Example}

\allowdisplaybreaks

\begin{document}

\title{Multi-Agent Distributed Optimization With \\ Feasible Set Privacy\thanks{This work was presented in part at CISS 2024.}}

\author{Shreya Meel \qquad Sennur Ulukus\\
\normalsize Department of Electrical and Computer Engineering\\
\normalsize University of Maryland, College Park, MD 20742 \\
\normalsize {\it smeel@umd.edu} \qquad {\it ulukus@umd.edu}}

\date{}

\maketitle

\vspace*{-0.8cm}

\begin{abstract}
We consider the problem of decentralized constrained optimization with multiple agents $E_1,\ldots,E_N$ who jointly wish to learn the optimal solution set while keeping their feasible sets $\mathcal{P}_1,\ldots,\mathcal{P}_N$ private from each other. We assume that the objective function $f$ is known to all agents and each feasible set is a collection of points from a universal alphabet $\mathcal{P}_{alph}$. A designated agent (leader) starts the communication with the remaining (non-leader) agents, and is the first to retrieve the solution set. The leader searches for the solution by sending queries to and receiving answers from the non-leaders, such that the information on the individual feasible sets revealed to the leader should be no more than nominal, i.e., what is revealed from learning the solution set alone. We develop achievable schemes for obtaining the solution set at nominal information leakage, and characterize their communication costs under two communication setups between agents. In this work, we focus on two kinds of network setups: i) ring, where each agent communicates with two adjacent agents, and ii) star, where only the leader communicates with the remaining agents. We show that, if the leader first learns the joint feasible set through an existing private set intersection (PSI) protocol and then deduces the solution set, the information leaked to the leader is greater than nominal. Moreover, we draw connection of our schemes to threshold PSI (ThPSI), which is a PSI-variant where the intersection is revealed only when its cardinality is larger than a threshold value. Finally, for various realizations of $f$ mapped uniformly at random to a fixed range of values, our schemes are more communication-efficient with a high probability compared to retrieving the entire feasible set through PSI.
\end{abstract}

\newpage

\section{Introduction}\label{intro}
\begin{figure}[t]
    \centering
    \includegraphics[width=0.5\textwidth]{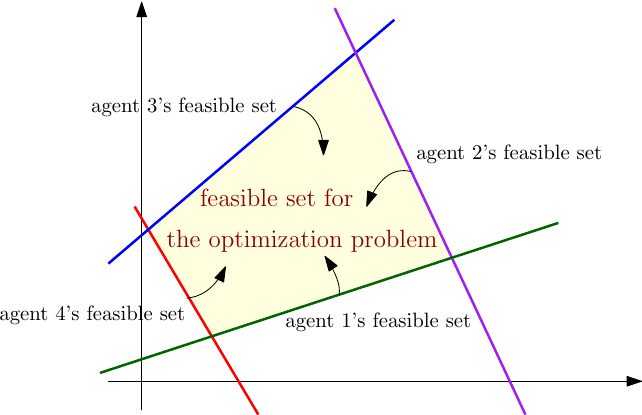}
    \centering
    \caption{Feasible set for a distributed optimization problem with four agents each with a linear inequality constraint.}
    \label{visualize}
\end{figure}

%privacy risks in dist opt
In distributed optimization, agents collaborate to find the optimal solution of a global objective function. Each agent has its own feasible set and the solution of the constrained optimization problem should be present in the intersection of all feasible sets as shown in Fig.~\ref{visualize}. The feasible set contains sensitive information of an agent, and should be kept private. For instance, consider the problem of allocating charging schedules to electric vehicles (EVs) \cite{pappasDCOP}, to minimize load fluctuations on a power grid subject to maximum charge and energy constraints for each EV. Then, the maximum charge for an EV being zero suggests that the owner of the EV (agent) is away, leaking private information about the agent. Most importantly, the iterative optimization algorithms \cite{asumanDCOP} involve exchanging solution estimates among agents over multiple rounds which, on collusion of all but one agent reveals information on the target agent's feasible set, breaching its privacy. 

% DP and homomorphic encryption solutions
As a solution, differential privacy (DP) \cite{dworkDP} based optimization algorithms alleviate this problem to some extent by adding noise to the estimates before exchanging them. For instance, \cite{pappasDCOP} incorporates DP in their algorithm to protect the privacy of agents' information in their respective constraints. This procedure retains some uncertainty on the agents' feasible sets  because of the noise-added estimates observed by the participating agents. Besides feasible sets, iterative distributed algorithms risk the privacy of agents' local functions as well. The global objective is given by the sum of these local functions and guaranteeing DP to agents in these scenarios was studied in \cite{vaidyaPDOP, diggaviDCOP, vaidyaDCOP,dingPDOP}. DP-based guarantees to protect user information in the objective function and constraints of a linear program was studied in \cite{hsuPSLP}. However, the main disadvantage of DP-based approaches is that they compromise the accuracy of solutions owing to the noise added. Moreover, they allow some certainty in an adversary's guess of the agent's private information, unlike full uncertainty as guaranteed by information-theoretic privacy. To achieve accurate solution, homomorphic encryption schemes can be incorporated in the distributed algorithm as in \cite{luhomomorphicPDOP}. However, the privacy in such schemes is provided by hardness in computing certain functions, which can be broken through infinite computing power.

% PSI as a step
To the best of our knowledge, none of the existing works address the problem of guaranteeing privacy to user's constraints in an information-theoretic sense. In this regard, we formulate distributed optimization with feasible set privacy (DOFSP), where the goal of the agents is to solve an optimization problem while keeping their feasible sets information-theoretically private from each other. For instance, in Fig.~\ref{visualize}, the agents describe their feasible sets by a set of discrete points satisfying the linear inequality constraints. The feasible set of each agent is a finite list of values from a universal alphabet and the objective function $f$ is known to all agents. The agents do not collude among themselves, and are assumed to be \emph{honest but curious}, in the sense that they follow the communication protocol truthfully, but are curious to learn information on each other's private feasible sets. To search for the optimum set $\mathcal{P}^*$, one of the agents (\emph{leader}) privately learns whether the best solutions in its own feasible set are also present in the feasible sets of the remaining (\emph{non-leader}) agents, while learning \emph{no further information} on their individual feasible sets. In fact, the agents learn no more information on the intersection of the feasible sets, than what is revealed from learning the solution set alone. We call this information leakage as \emph{nominal information leakage}. 

%PSI using PIR
The agents achieve this by first mapping the feasible sets into incidence vectors \cite{zhushengPSI, zhushengMultiPSI}, storing them on non-colluding databases and exchanging carefully designed messages among themselves till the optimum solution set $\mathcal{P}^*$ is learned  by a designated agent, i.e., leader. Our protocols require the incidence vectors of feasible sets to be replicatedly stored in at least two non-colluding databases of the non-leader agents. This is along the lines of the information-theoretic formulation of private information retrieval (PIR) \cite{jafarPIR, banawanPIRcoded, chaotian, ulukusPIRLC, sajaniPIRmagazine} where messages are replicated across multiple databases. In PIR, a user wants to retrieve one of these messages, without revealing the identity of the message to an individual server. Specifically, our protocols are derived from symmetric PIR (SPIR) schemes \cite{jafarSPIR}. In SPIR, the additional requirement beyond PIR is that the messages in the databases, other than the desired one, should be hidden from the retrieving user.

\begin{figure}[t]
    \centering
    \begin{subfigure}[t]{0.4\textwidth}
        \centering
        \includegraphics[height=\textwidth]{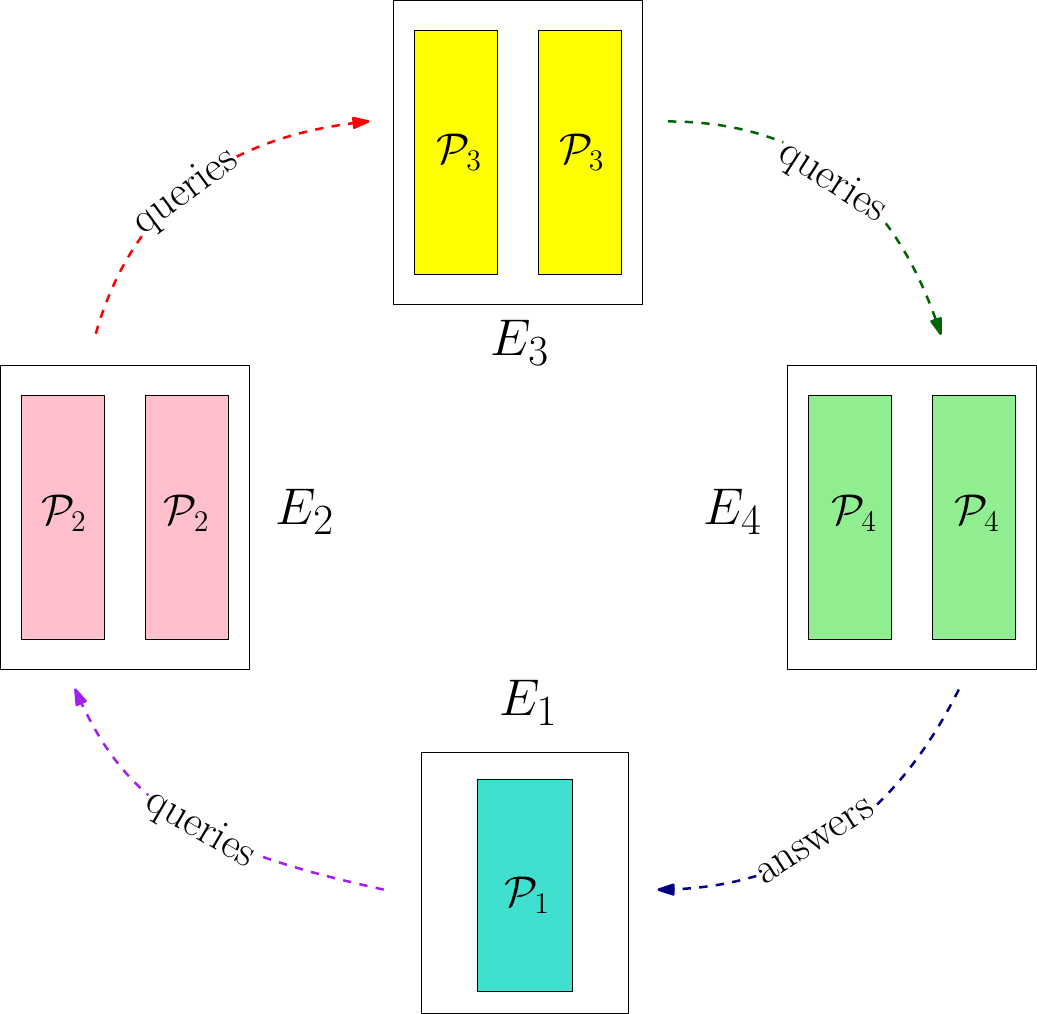}
        \caption{Ring setup.}
    \end{subfigure}
    \qquad \qquad 
    \begin{subfigure}[t]{0.42\textwidth}
        \centering
        \includegraphics[height=\textwidth]{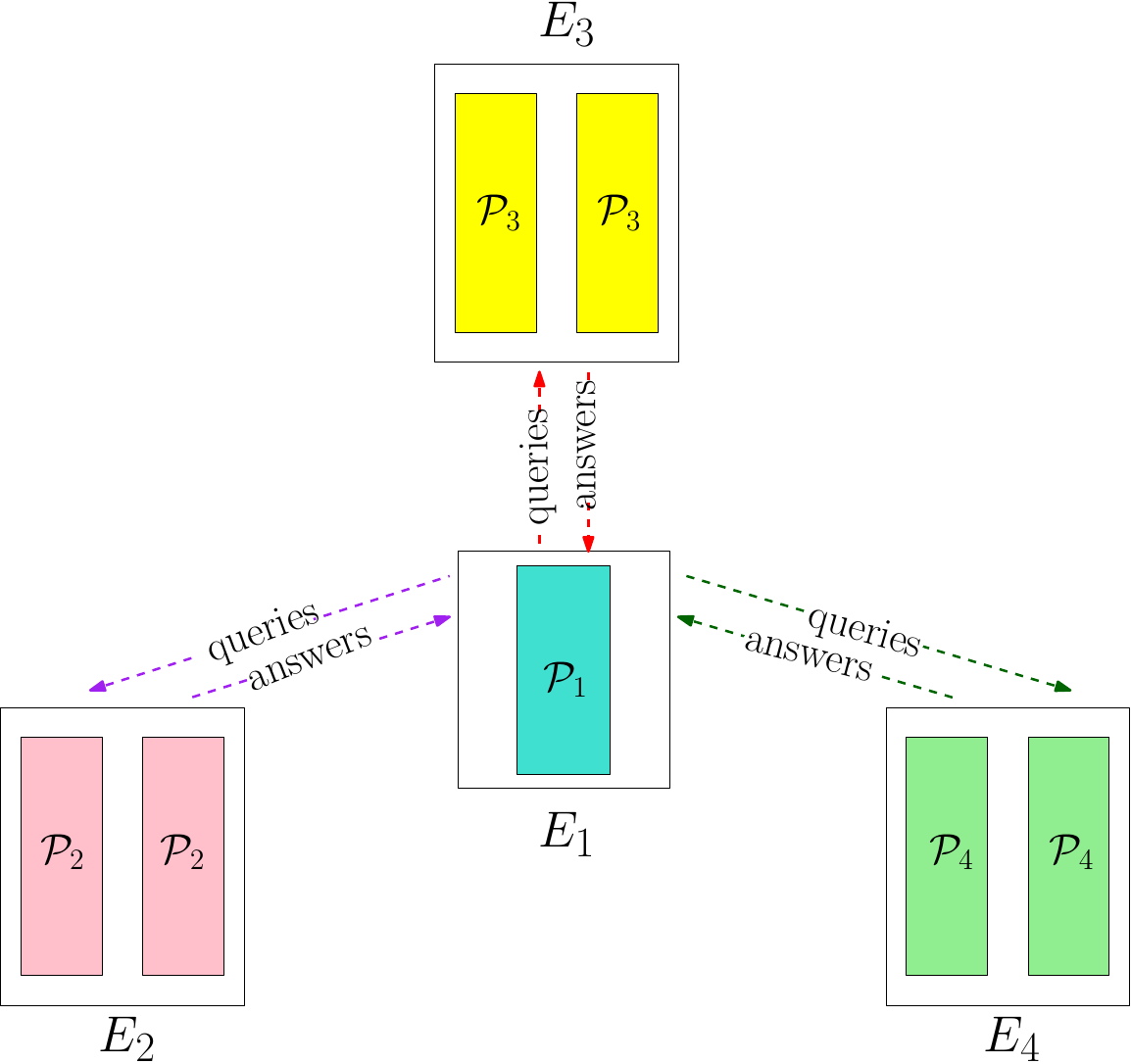}
        \caption{Star setup.}
    \end{subfigure}
    \caption{Two types of communication setups for $N=4$ entities. Leader entity $E_1$ has a single database, each non-leader entity, $E_2$, $E_3$, $E_4$, has $N_i=2$ databases, $i=2, 3, 4$.}
    \label{comm_setups}
\end{figure}

% Case 1: N=2 (method)
For the simple two-agent case, our scheme proceeds as follows. The leader privately checks if the \emph{best} solutions in the leader's set also appear in the other agent's set, by invoking the cardinality PSI (CarPSI) protocol which returns the number of such elements. If none of the candidate elements are present in the intersection, i.e., CarPSI returns a zero, the agent re-invokes CarPSI, this time with the \emph{next best} set of elements. Otherwise, the solution appears on both feasible sets, hence at the intersection, giving the optimum function value and the leader learns the number of such elements. Next, the leader finds $\mathcal{P}^*$ by executing FindPSI protocol, which privately finds the set intersection, with the knowledge of its cardinality. Alternatively, one can view the DOFSP scheme as a sequential application of a two-user \emph{threshold private set intersection} (ThPSI) protocol wherein the set intersection is evaluated only if its cardinality is greater than or equal to a threshold, while revealing no further information about the sets as in PSI \cite{tsudikPSI}. As an example, an important instance of ThPSI arises in a ride sharing app where each passenger wants to share information on their trajectory only when there is sufficient overlap between the routes. In the context of our scheme, the ThPSI operates on subsets dictated by the function values and the threshold is 1 in our case. Hence, we obtain a ThPSI protocol guaranteeing perfect information-theoretic privacy unlike the existing protocols of \cite{zhangthPSI, ghoshthPSI} which guarantee computational privacy only. 

%Case 2: N>2 (method)
Next, we extend our scheme to the multi-agent scenario. In terms of design of the communication protocol, we consider two setups as shown in Fig.~\ref{comm_setups}:
\begin{enumerate}
    \item \emph{Ring}, where each agent communicates with its two neighboring agents only. The leader sends queries to the next agent, who after processing them, sends queries to the next agent, and so on. The last agent sends answers back to the leader.
    \item \emph{Star}, where the leader communicates with the remaining $N-1$ agents that do not communicate among themselves. The leader sends queries to the databases of the remaining agents, who respond to it with their answers.
\end{enumerate}

We provide a DOFSP scheme for the ring setup by extending CarPSI and FindPSI to primitives CarPSI-ring and FindPSI-ring, respectively. The CarPSI-ring primitive starts with the leader sending shares \cite{shamir} of its incidence vector to two non-colluding databases of the neighboring agent. Each database then uses the received message and the stored incidence vector and computes a message to be sent over the database of the next agent and so on. Finally, the leader collects the messages from the $N$th database pair and computes the cardinality of the set intersection. If this is greater than or equal to the threshold value, the leader retrieves the set intersection by retrieving two additional messages from the $N$th agent by executing FindPSI-ring. For the star setup, our DOFSP scheme is a sequential application of the multi-party PSI scheme of \cite{zhushengMultiPSI}.

%Summary of results
The main contributions of this paper can be summarized as follows:
\begin{itemize}
    \item We propose achievable DOFSP schemes for the two-user and multi-user setups, where we deal with two communication setups, ring and star in the latter case, and characterize their communication efficiency in terms of the upload and download costs, which are the number of symbols exchanged between the agents.
    \item We show that, unlike our DOFSP schemes, the naive approach of learning the feasible set first, and then evaluating $\mathcal{P}^*$ leaks more information than the nominal and incurs a higher or equal communication cost.
    \item We numerically evaluate the probability of the communication cost of our DOFSP scheme being equal to that of the naive scheme, and show that this probability is small. Specifically, under uniform mapping of $f$, we evaluate this probability for both the setups with $N\geq 2$ with a fixed feasible set assignment, and show through our numerical results that this probability is low.
\end{itemize}

The rest of the paper is organized as follows. Section \ref{problemForm} describes the DOFSP problem setting and Section \ref{results} states the results of the paper. In Section \ref{achievable2}, we present our DOFSP scheme with $N=2$ agents. In Sections \ref{achievable>2_ring} and \ref{achievable>2_star}, we present the DOFSP schemes under the ring and star setups, respectively, for $N>2$. Finally, we conclude the paper in Section \ref{sec:conclude}.

{\bf Notations:} To denote the set of integers $\{1,2,\ldots,n\}$ for a positive integer $n$, we use the notation $[n]$. For positive integers $m,n$ with $m<n$, the notation $[m:n]$ represents the set $\{m,m+1,\ldots,n-1,n\}$. We use the symbol $\mathbb{N}$ to denote the set of positive integers. 

\section{Problem Formulation}\label{problemForm}
\subsection{General Problem Setting}
We consider a setting with $N$ entities (agents) who jointly seek the optimal solution (or optimal solutions) of an objective function, subject to satisfying the constraints of all participating entities. Each entity $E_i, \, i\in[N]$, has a feasible set $\mathcal{P}_i$ with elements from a finite indexed alphabet set $\mathcal{P}_{alph}$ of size $|\mathcal{P}_{alph}|=K$. Therefore, each $\mathcal{P}_i$ of size $P_i$ is a subset of $\mathcal{P}_{alph}$ and is private to $E_i$. For the existence of a non-empty solution set, we assume the feasible sets to have a non-empty intersection, i.e., $\mathcal{P}:=\mathcal{P}_1\cap \mathcal{P}_2\cap\ldots\cap\mathcal{P}_N\neq \emptyset$. 

Each $E_i$ casts its feasible set $\mathcal{P}_i$ into an \emph{incidence vector}, which is a $K$-length binary column vector $X_i$ indicating the presence (or absence) of an element by $1$ (or $0$), respectively. We write the $k$th entry, for $k\in[K]$, of $X_i$ as follows,
\begin{align}
     X_i(k) =\begin{cases} 
      1, & \text{if }\mathcal{P}_{alph}(k) \in \mathcal{P}_i, \\
      0, & \text{otherwise}, 
   \end{cases} 
\end{align}
where $\mathcal{P}_{alph}(k)$ is the element of $\mathcal{P}_{alph}$ at index $k$. Each entity $E_i$ has access to $N_i\geq 2$ non-colluding databases, where each database of $E_i$ stores the incidence vector $X_i$. The numbers $P_i$ and $N_i$ are known to all the entities for each $i\in [N]$. 

The objective function $f$ is modeled as a random mapping $f:\mathcal{P}_{alph}\to \mathcal{T}$ independent of the sets $\mathcal{P}_i$, where the set $\mathcal{T}$ takes $\tau$ distinct values. For our subsequent analysis, we assume that $f$ maps each element in $\mathcal{P}_{alph}$ uniformly to an element in $\mathcal{\tau}$. In other words, for any $x \in \mathcal{P}_{alph}$ and  $y \in \mathcal{T}$, $\mathbb{P}(f(x)=y)=\frac{1}{\tau}$. The mapping is done independently for each $x$.
 
The goal of $E_1,\ldots,E_N$ is to find the solution set $\mathcal{P}^*$ where $f$ attains the optimum (minimum or maximum) value within the overall feasible set $\mathcal{P}$, while keeping their individual feasible sets private from each other. That is, for a minimization problem, the entities follow a communication protocol to find the solution to:
\begin{align}
    &\underset{x}{\text{minimize}} \quad \ f(x) \nonumber\\
    &\text{subject to} \quad x\in \mathcal{P}.
\end{align}
However, the protocol should not reveal to $E_i$ any information about the feasible sets other than its own, i.e., $\mathcal{P}_i$ beyond what is implied by $\mathcal{P}^*$. Thus, the associated information leakage on the remaining feasible sets to every agent should be \emph{nominal} (minimal) which we define as follows.

\begin{definition}[Nominal information leakage]
     Nominal information leakage is the amount of information on the other feasible sets revealed to a specific entity by learning the solution set $\mathcal{P}^*$ only, given its own feasible set and the knowledge of $f$. For $E_i, \,i\in[N]$, it is denoted by $\mathcal{I}_{nom,i}$ and defined as
    \begin{align}\label{eq_nom_info_leak}
        \mathcal{I}_{nom,i} = I(\mathcal{P}^*;\mathcal{P}_1, \ldots, \mathcal{P}_{i-1}, \mathcal{P}_{i+1}, \ldots, \mathcal{P}_N|\mathcal{P}_i,f).
    \end{align}
\end{definition}

We assume that the communication protocol is started by a designated entity, \emph{leader} (say, $E_1$). It is also the first entity that learns $\mathcal{P}^*$ and communicates it to the other entities. Given a realization of $f$, we define two vectors: Global function profile $\bm{\mu}$ and local function profile $\bm{\alpha}$ which indicate the number of elements in $\mathcal{P}_{alph}$ and $\mathcal{P}_1$, respectively, that map to the same function value via $f$.
\begin{enumerate}
    \item{\textbf{Global function profile:}}  Let $f_1,f_2,\ldots ,f_T$ where $T\in [\tau]$ be ordered such that $f_1$ is the best and $f_T$ is the worst function value in terms of the optimization objective. Then, $\bm{\mu}=[\mu_1, \mu_2,\ldots, \mu_T]$, where $\mu_1+\mu_2 + \ldots + \mu_T= K$,  $\mu_t \geq 1$,  $t\in [T]$. All entities have the knowledge of $\bm{\mu}$.
    \item{\textbf{Local function profile:}} Let the equi-cost elements in $\mathcal{P}_1$ be ordered from best to worst as $f_{i_1},f_{i_2}, \ldots, f_{i_L}$ where $ i_1<i_2<\ldots<i_L$. Further, let $\mathcal{J}_r$ be the set of indices in $X_1$ corresponding to the $r$th best solution in $\mathcal{P}_1$, with  $f_{i_r}$ as the respective function value. Then, $\bm{\alpha}=[\alpha_1, \alpha_2, \ldots, \alpha_L]$ with $\alpha_r$ denoting the multiplicity of the $r$th best solution, i.e., $|\mathcal{J}_r|=\alpha_r$ for all $r\in[L]$. Each $\alpha_r\geq 1$ and $\sum_{r=1}^L\alpha_r = P_1$. Only $E_1$ has the knowledge of $\mathcal{J}_{1:L}$ and $\bm{\alpha}$.
\end{enumerate}

Let $\mathcal{M}'_{in}$ and $\mathcal{M}'_{out}$ denote the incoming and outgoing messages of $E_1$ in the communication protocol. Then, to ensure the correct recovery of $\mathcal{P}^*$ at $E_1$, we must have,
\begin{align}\label{eq:reliab}
    [\hbox{reliability at } E_1] \qquad H(\mathcal{P}^*|\mathcal{M}'_{out}, \mathcal{M}'_{in},\mathcal{P}_1,f)=0.
\end{align}
The leader $E_1$ should not gain, from the messages it sends and receives, any information on $\mathcal{P}_i, i\neq 1$, beyond the nominal leakage,
\begin{align}\label{eq:leakage_leader}
     [\hbox{privacy of } E_{2:N} \hbox{ against } E_1] \qquad I(\mathcal{M}'_{in},\mathcal{M}'_{out};\mathcal{P}_{2:N}|\mathcal{P}_1,f)=\mathcal{I}_{nom,1}.
\end{align}
To ensure the privacy of the outgoing messages, the databases of each non-leader $E_i$, $i\in \{2,\ldots,N\}$, share a set of common randomness symbols, denoted by $\mathcal{S}_i$. Hence, the content stored at each database of $E_i$ is $X_i$ and $\mathcal{S}_i$. Let $\mathcal{M}_{in,i,j}$ and $\mathcal{M}_{out,i,j}$ denote the incoming and outgoing messages of database $j$ of $E_i$. Then, the database learns no information about the remaining feasible sets, $\mathcal{P}_{[N]\setminus i}$, 
\begin{align}\label{eq:leakage_non-leader}
    [\hbox{privacy of } E_{[N]\setminus i} \hbox{ against database $j$ of } E_i] \qquad I(\mathcal{M}_{in,i,j},\mathcal{M}_{out,i,j};\mathcal{P}_{[N]\setminus i}|\mathcal{P}_i,\mathcal{S}_i)=0.
\end{align}
In our communication protocol, the entities other than $E_1$ do not directly participate, and only their databases are involved in the respective communications. Therefore, it suffices if the individual databases do not learn any information about the other feasible sets. This ensures that the information leaked to any non-leader $E_i$ is simply $\mathcal{I}_{nom,i}$.

An achievable scheme for distributed optimization with feasible set privacy (DOFSP), under a ring or star setup, should satisfy the reliability and privacy requirements in \eqref{eq:reliab}, \eqref{eq:leakage_leader} and \eqref{eq:leakage_non-leader}. Next, we specify the messages sent and received in the protocol for each communication setup.

\subsection{Ring Setup}
Each non-leader entity $E_i$ has $N_i\geq 2$ non-colluding databases. Let them be labeled as $1,2,\ldots,N_i, \forall i$. We assume that there exists a noiseless, secure communication link between database $j$ of $E_i$ with database $j$ of $E_{i+1}$, for all $i\in [2:N-1]$. There also exist noiseless, secure links between $E_1$ and the databases of $E_2$ and $E_N$. Without loss of generality, assume that databases in the set $\mathcal{N}_{eff} = \{1,\ldots, N_{eff}\}$ where $N_{eff} = \min_{i\in [2:N]} N_i$ form a secure ring across the entities, with $N_{eff}\geq 2$. Alternatively, only the databases in $\mathcal{N}_{eff}$ of every non-leader entity participate in the communication protocol.

The communication flow is as follows. $E_1$ sends queries $Q_{1,j}$ to database $j\in\mathcal{N}_{eff}$ of $E_2$. For the intermediate entities, $E_i, i\in [2:N-1]$, each database $j$ of $E_i$ sends a query $Q_{i,j}$ to database $j$ of $E_{i+1}$. Finally, database $j$ of $E_N$ sends its answer $A_j$ to $E_1$. Thus, the communication flow can be described by a ring. The corresponding messages exchanged are
\begin{align}
    \mathcal{M}'_{out}=Q_{1,j}, \quad \mathcal{M}'_{in}=A_{j}, \quad j\in \mathcal{N}_{eff}.
\end{align}
Further, for the databases of $E_i,i\in[2:N-1]$, we have
\begin{align}
    \mathcal{M}'_{out,i,j}=Q_{i,j}, \quad \mathcal{M}'_{in,i,j}=Q_{i-1,j}, \quad j\in \mathcal{N}_{eff},
\end{align}
and for $E_N$, we have
\begin{align}
    \mathcal{M}'_{out,N,j}=A_j, \quad \mathcal{M}'_{in,N,j}=Q_{N-1,j}, \quad j\in \mathcal{N}_{eff}.
\end{align}
Each $Q_{i,j}$ is a function of $Q_{i-1,j}$, $X_i$ and $\mathcal{S}_i$ and each $A_{j}$ is a function of $Q_{N-1,j}, X_N$ and $\mathcal{S}_N$. The communication cost $C_{ring}$ is the total number of symbols transmitted between entities till the protocol is completed.

\subsection{Star Setup} 
In the star setup, we adopt the SPIR framework for PSI as in \cite{zhushengPSI, zhushengMultiPSI}. Entity $E_i$ has $N_i$ databases where $X_i$ is stored. The index $l^*\in [N]$ of the leader is assigned as
\begin{align}\label{find_lstar}
    l^*=\argmin _{l\in [N]}\sum_{i=1, i\neq l}^N \bigg\lceil\frac{P_lN_i}{N_i-1}\bigg\rceil,
\end{align}
in order to minimize the communication cost of PSI. We relabel the entities to let $l^*=1$. $E_1$ has a secure, noiseless communication link with each individual database of the non-leader entities. Apart from sharing the common randomness $\mathcal{S}_i$, the databases of $E_i, i\in [2:N]$, do not collude among themselves. In the communication protocol, for $N\geq 2$, $E_1$  sends queries to and collects answers from the databases of the remaining entities. In particular, we have
\begin{align}
    \mathcal{M}'_{out}=\bigcup_{i\in[N]}\bigcup_{j\in [N_i]} Q_{i,j}, \quad \mathcal{M}'_{in}=\bigcup_{i\in[N]}\bigcup_{j\in [N_i]} A_{i,j}, \quad i\in\{2,\ldots,N\},
\end{align}
where $Q_{i,j}$ is the query sent to and $A_{i,j}$ is the answer received from the $j$th database of $E_i$. Similarly,
\begin{align}
    \mathcal{M}_{in,i,j}=A_{i,j}, \quad \mathcal{M}_{out,i,j}=Q_{i,j},
\end{align}
where each answer $A_{i,j}$ is a function of $Q_{i,j}$, $X_i$ and $\mathcal{S}_i$. (For $N=2$, we simplify $Q_{i,j}$ and $A_{i,j}$ to $Q_j$ and $A_j$, respectively.) The download cost $D_{star}$ refers to the total number of symbols of the operating field received by $E_1$ from all the non-leader databases. The upload cost $U_{star}$ is the number of symbols sent by $E_1$ to these databases from start to end of the protocol. Their sum $C_{star}=D_{star}+U_{star}$ reflects the total communication cost. If $N=2$, we denote these costs by $C$, $D$ and $U$, respectively. 

\section{Main Results} \label{results}
\begin{theorem}\label{thm1}
    For a $N=2$ agent DOFSP problem, given the feasible sets $\mathcal{P}_1$ and $\mathcal{P}_2$, the worst-case download cost $D$ depends on the position $R$ of the best function value in $X_1$ and is given by,
    \begin{align}\label{dlcost1}
        D&=
        \begin{cases}
         \big\lceil\frac{(R+\alpha_R-1)N_2}{N_2-1}\big\rceil, &   \alpha_R > X_{\mathcal{J}_R}^TX_2, \\
         \big\lceil\frac{RN_2}{N_2-1}\big\rceil, &  \alpha_R = X_{\mathcal{J}_R}^TX_2,
        \end{cases}
    \end{align}
    where $X_{\mathcal{J}_r}=\sum_{j\in\mathcal{J}_r }{e}_j$ and  ${e}_j$ is the $K$-length unit column vector with $1$ at the $j$th position. The upload cost $U$ for the above scheme is $KD$ symbols, resulting in $C=U+D=(K+1)D$. 
\end{theorem}

The proof of Theorem \ref{thm1} follows from our proposed scheme in Section \ref{achievable2}.

\begin{theorem}\label{thm2}
    For an $N > 2$ agent DOFSP problem under the ring setup, given the feasible sets $\mathcal{P}_1, \mathcal{P}_2, \ldots, \mathcal{P}_N$, the worst-case communication cost $C_{ring}$ is given by,
    \begin{align}
        C_{ring}=
        \begin{cases}
            2N\left(\sum_{r=1}^R\mu_r\right)+2\left(R-\sum_{r=1}^{R-1}\mu_r\right), & R<T,\\
            2NK+2\left(R-1-\sum_{r=1}^{R-1}\mu_r\right), &  R=T,
        \end{cases}
    \end{align}
    where $R\in [T]$ corresponds to the index of the optimum function value in $\bm{\mu}$. 
\end{theorem}

\begin{theorem}\label{thm3}
    For an $N>2$ agent DOFSP problem under the star setup, given the feasible sets $\mathcal{P}_1, \mathcal{P}_2, \ldots, \mathcal{P}_N$, the worst case download cost $D_{star}$ is given by,
    \begin{align}
        D_{star}=\sum_{i=2}^N \Big\lceil\frac{(\sum_{r=1}^R \alpha_r) N_i}{N_i-1}\Big\rceil,
    \end{align}
    where $R\in [L]$ corresponds to the index of the optimum function value in $\bm{\alpha}$. The upload cost $U_{star}$ for the above scheme is $KD_{star}$, giving $C_{star}=(K+1)D_{star}$.
\end{theorem}

The proofs of Theorems \ref{thm2} and \ref{thm3} follow from the schemes in Sections \ref{achievable>2_ring} and \ref{achievable>2_star}, respectively.

\section{The Case of $N=2$ Entities}\label{achievable2}
When $N=2$, we refer to the leader $E_1$ as \emph{client} and $E_2$ as \emph{server} and adopt a star communication setup (letting $\lceil\frac{P_1N_2}{N_2-1}\rceil \leq \lceil\frac{P_2N_1}{N_1-1}\rceil $). Note that, if $E_1$ obtains the joint feasible set $\mathcal{P}$ first using \cite{zhushengPSI}, and then finds $\mathcal{P}^*$ by comparing the function values, the associated download cost is $D_{PSI}=D_{PSI}(P_1,N_2)=\lceil{\frac{P_1N_2}{N_2-1}}\rceil$. 

In contrast, in our achievable scheme to seek for the optimal solution set, $E_1$ sequentially checks the presence of the set of elements that achieve the best function value. The answers received in the current iteration, after being decoded, decide the set of queries for the next iteration. We present an example before formally describing our scheme.

\begin{example}\label{motivex}
    Consider the following scenario. A universal movie-rating system provides a score between $1$ to $5$ to the movies, with $5$ being the best. Let $\mathcal{P}_{alph}=\{A, B, C, D, E, F, G, H\}$ be the collection of movie IDs. Entities $E_1$ and $E_2$ have subsets of these IDs as their feasible sets as
    \begin{align}
        \mathcal{P}_1=\{A, C, D, G\}, \qquad \mathcal{P}_2=\{B, C, D, G, H\}, 
    \end{align}
    and therefore the incidence vectors as
    \begin{align}
        X_1 = [1 \ 0 \ 1 \ 1 \ 0 \ 0 \ 1 \, 0]^T, \qquad X_2 = [0 \ 1 \ 1 \ 1 \ 0 \ 0 \ 1 \ 1]^T.
    \end{align}
    The goal of the entities is to find the set of movies with the highest score present in the sets of both entities. Let the scores be given by
    \begin{align}
        &f(B)=f(E)=f(H)=5, \nonumber\\
        &f(A)=f(C)=f(G)=4, \nonumber\\
        &f(D)=3, \nonumber\\
        &f(F)=2.
    \end{align} 
    Therefore, we have $\bm{\alpha} = [3,1]$. For this problem, $\mathcal{P}^*$ is $\{C,G\}$.
    
    First, $E_1$ checks for the presence of $\{A\}$, $\{C\}$ or $\{G\}$ (the elements that attain the maximum score of $4$ in $\mathcal{P}_1$) in $\mathcal{P}_2$, i.e., for $\mathcal{J}_1=\{1,3,7\}$. To do so, $E_1$ picks a random vector ${h}_1$ from $\mathbb{F}_q^8$ with $q>3$ and prime, and sends the queries, 
    \begin{align}
        Q_1 = {h}_1, \qquad Q_2 = {h}_1 + {e}_1 + {e}_3 + {e}_7,
    \end{align}
    to the first two databases of $E_2$. We assume that the databases of $E_2$ share a common random variable $S_1\in \mathbb{F}_q^8$. The answers returned by the corresponding databases are,
    \begin{align}
        A_1 = Q_1^TX_2 + S_1, \qquad A_2 = Q_2^TX_2 +S_1.
    \end{align}
    Then, entity $E_1$ computes $A_2-A_1=2$ and concludes that there are $2$ elements with score $4$ in $\mathcal{P}_1\cap \mathcal{P}_2$. In the first phase, $E_1$ downloads $2$ symbols, one from each database. 
    
    Now, to find which elements out of $\{A,C,G\}$ form the solution set, $E_1$ checks for the presence of any two $\{A,C\}$ (and $\{C,G\}$ and $\{A,G\}$) in $\mathcal{P}_2$. Based on the number of databases $N_2$ at $E_2$'s side, the queries will differ for this phase as tabulated in Table~\ref{tabex}. For instance, in the case of $N_2=3$, the answers downloaded by $E_1$ are $({h}_1+{e}_1)^TX_2+S_1$ from database 3, ${h}_2^TX_2+S_2$ from database 1, and  $({h}_2+{e}_3)^TX_2+S_2$ from database 2, where the databases share $S_1,S_2\in \mathbb{F}_q^8$ as common random variables. In this phase, the number of downloaded symbols is $4$, $3$ and $2$ if $N_2$ is $2$, $3$ or $4$, respectively. If there are more than $4$ databases then queries in the last row are sent to any two of databases 3 to $N_2$.  
    
    \begin{table}[t]
    \centering
        \begin{tabular}{|c|c|c|c|c|}
        \hline
          $N_2$ & database 1 & database 2 & database 3 & database 4\\
          \hline
          2 & ${h}_2$, ${h}_3$ & ${h}_2+{e}_1$, ${h}_3+{e}_3$ & - & -\\
          \hline
          3 & ${h}_2$ & ${h}_2+{e}_3$ & ${h}_1+{e}_1$ & -\\
          \hline
          4 & & & ${h}_1 + {e}_1$ & ${h}_1+{e}_3$\\
          \hline
        \end{tabular} 
        \caption{Queries sent to check the presence of $\{A,C\}$ in $\mathcal{P}_2$.}
        \label{tabex}
    \end{table}
    
    Now, with the same feasible sets consider a different score mapping where $f(C)=f(G)=3$ and the rest of the scores are the same as before, i.e.,
    \begin{align}
        &f(B)=f(E)=f(H)=5, \nonumber\\
        &f(A)=4, \nonumber\\
        &f(C)=f(D)=f(G)=3, \nonumber\\
        &f(F)=2.
    \end{align} 
    Therefore, we have $\bm{\alpha}=[1, 3]$. Entity $E_1$ checks for the existence of $\{A\}$ in $E_2$ by sending queries ${h}_1$ and ${h}_1+{e}_1$ to two databases of $E_2$. Since the answers from $E_2$ reveal the absence of $A$ in $\mathcal{P}_2$, $E_1$ moves to the next-best value $3$ and checks for the existence of $\{C\}$, $\{D\}$ or $\{G\}$. On learning that all three elements are common in $\mathcal{P}_2$, $E_1$ directly learns that the optimal solution is $\{C,D,G\}$. The download cost in this case is $4$ if $N_2=2$ and $2$ if $N_2>3$. 
    
    Next, consider yet another score mapping where $f(G)=5$, and $f(x)=4$, otherwise, i.e.,
    \begin{align}
        &f(G)=5, \nonumber\\
        &f(A)=f(B)=f(C)=f(D)=f(E)=f(F)=f(H)=4. 
    \end{align} 
    Therefore, we have $\bm{\alpha}=[1, 3]$. This time, $E_1$ starts by checking the existence of $G$ by sending queries ${h}_1$ and ${h}_1+{e}_7$ to two databases with $E_2$. By downloading $2$ symbols as answer, $E_1$ finds the optimal solution to be $\{G\}$.
\end{example}

\subsection{Proposed Scheme}
As illustrated in the example, our scheme consists of two primitives: 1) CarPSI and 2) FindPSI. Let $M_{i_r}:=X_{\mathcal{J}_r}^TX_2$ denote the multiplicity of elements with function value $f_{i_r}$ in $\mathcal{P}_1\cap \mathcal{P}_2$. Before the start of communication, both entities agree on a finite field $\mathbb{F}_q$ with a prime $q$ which satisfies $q > \max_{r\in[L]}\alpha_r$. Now, we describe the functionalities CarPSI and FindPSI for any $r\in [L]$.

\subsubsection{CarPSI($X_{\mathcal{J}_r}, N_2$)}\label{sec:CarPSI_N=2}  
CarPSI($X_{\mathcal{J}_r}, N_2$) computes the cardinality $M_{i_r}$ for which $E_1$ evaluates the dot product of $X_{\mathcal{J}_r}$ with $X_2$ without revealing $\mathcal{J}_r$ to any database of $E_2$. Further, CarPSI should not reveal any additional information on $X_2$ beyond $M_{i_r}$. Depending on $r$, there can be two cases. 

\underline{Case i)}: If $r=(k-1)(N_2-1)+ 1$ for some $k\in \mathbb{N}$, $E_1$  uniformly  selects $K$ elements $h_k(j), j\in[K]$ independently from $\mathbb{F}_q$ to form the vector ${h}_k =[h_k(1), h_k(2), \ldots, h_k(K)]^T$.  Now, $E_1$ randomly selects 2 databases (say the first 2) from the $N_2$ databases available to $E_2$ to participate in the protocol. The queries sent are
\begin{align}
    Q_1 = {h}_k,  \qquad Q_2 = {h}_k+ X_{\mathcal{J}_r}.\label{query2}
\end{align}
To generate answers, the databases calculate the dot product $Q_n^TX_2$, $n=1,2$ and add a new symbol to $S_k\in \mathcal{S}$,
\begin{align}
    A_1 &= Q_1^T X_2 +S_k=  \sum_{j=1}^K h_k(j) X_2(j) + S_k,\\
    A_2 &= Q_2^T X_2 +S_k= \sum_{j=1}^K \big(h_k(j)  X_2(j) + X_{\mathcal{J}_r}(j) X_2(j)\big) +S_k.
\end{align}
All operations are performed in $\mathbb{F}_q$. 

\underline{Case ii)}: If $r=(k-1)(N_2-1)+l$, where $l\in \{2,\ldots,N_2-1\}$ for some $k\in \mathbb {N}$, the same vector $h_k$ used in CarPSI($X_{\mathcal{J}_{r-1}}, N_2$) is reused to form the query ${h}_k+ X_{\mathcal{J}_r}$. It is sent to one of the databases $n$ for which $h_k$ was not involved in the query. In response, database $n$ sends $A_n = Q_n^T X_2 +S_k$. Having received the answers, $E_1$ computes $M_{i_r}$ as,
\begin{align}
    M_{i_r} &= A_n - A_1 \\
    &= (Q_n - Q_1)^T X_2 = X_{\mathcal{J}_r}^T X_2.
\end{align}

\subsubsection{FindPSI($\mathcal{J}_r,M_{i_r},N_2$)} 
This functionality evaluates the private set intersection (PSI) between two entities, with the knowledge of the size of the intersection. Since $M_{i_r}$ is known, it suffices if $E_1$ leaves any single index (say, $j$) from $\mathcal{J}_r$ out, to learn the set intersection. Let $\mathcal{\Bar{J}}=\mathcal{J}_r\setminus \{j\}$ be the set on which PSI is to be executed, with $|\Bar{\mathcal{J}}|=\alpha_r-1$. Let $r=(k-1)(N_2-1)+l$ for some $k\in\mathbb{N}$ and $l\in [0:N_2-2]$, and $\alpha_r-1=(N_2-l-1)+ p(N_2-1) + s$, for some $p\in \mathbb{N}\cup \{0\}, s\in [0:N_2-2]$. Depending on $l$, there are two cases.

\underline{Case i)} If $l=0$, then $\alpha_r-1 = (p+1)(N_2-1)+s$. $E_1$ generates $p+2$ independent, uniformly random vectors ${h}_{k+1},\ldots, h_{k+p+2}\in \mathbb{F}_q^K$. 
 
\underline{Case ii)} If $l\in [N-2]$, $l+1$ databases have been previously sent queries to, using the same random vector $h_k$ in CarPSI. Therefore, $E_1$ reuses ${h}_k$ to submit queries to the unused (assume, $l+2,\ldots,N_2$) databases, and picks $p+1$ new random, independent vectors $h_{k+1},\ldots,h_{k+p+1}$ for the subsequent queries.

$E_1$ sends $Q_1=\{h_y, \ y\in [k+1:k+p+2]\}$ as query in case i) and $Q_1=\{h_y, \ y\in [k+1:k+p+1]\}$ as query in case ii) to database 1. To databases $n\in[2:N_2]$, the queries for the corresponding $y$ are $Q_n=h_y+ {e}_j, j\in \Bar{\mathcal{J}}$. In both cases, the answer returned corresponding to each $y$ is
\begin{align}
    A_n=Q_n^TX_2 +S_y, \quad n\in[N_2],
\end{align} 
where $S_y\in \mathcal{S}$. Formally, the integer $y$ ranges across $[k+1:k+p+2]$ if $n\in[s+1]$ and across $[k+1:k+p+1]$ otherwise, in case i). In case ii) if $s\leq l$, then $y\in [k+1:k+p+1]$ if $n\in[s+1]$, $y\in [k+1:k+p]$ if $n\in s+2:l+1$ and $y\in [k:k+p]$ if $n\in [l+2:N_2]$. The sub-case where $s>l$ can be similarly handled. To decode each element in $X_2(\Bar{\mathcal{J}})$, $E_1$ evaluates $A_n-A_1$. By doing so, $E_1$ learns $X_2(\mathcal{J}_r)$ since $M_{i_r}$ is known. 

\begin{algorithm}[t]
    \caption{Scheme to find $\mathcal{P}^*$}\label{alg1}
    \hspace*{\algorithmicindent} \textbf{Input:} $\mathcal{P}_1,f,\bm{\alpha},N_2$\\
    \hspace*{\algorithmicindent} \textbf{Output:} $\mathcal{P}^*$
    \begin{algorithmic}
    \State $L \gets \hbox{ length of } \bm{\alpha}$
    \For {$r \in \{1,2,\ldots, L\}$}
    \If {$r<L$ or $\alpha_L>1$}
    \State $M_{i_r} = \text{CarPSI}(X_{\mathcal{J}_r},N_2)$
    \If{$1\leq M_{i_r}\leq \alpha_r-1$}
    \State $\mathcal{P}^*$ = FindPSI($\mathcal{J}_{r},M_{i_r}, N_2$)
    \State \hspace*{\algorithmicindent} \textbf{break}
    \ElsIf {$M_{i_r}==\alpha_r$}
    \State $\mathcal{P}^*=\mathcal{P}_{alph}(\mathcal{J}_r)$
    \State \hspace*{\algorithmicindent} \textbf{break}
    \EndIf
    \Else 
    \State $\mathcal{P}^*=\mathcal{P}_{alph}(\mathcal{J}_r)$
    \EndIf
    \EndFor
    \State \Return $\mathcal{P}^*$
    \end{algorithmic}
\end{algorithm}

\subsubsection{Main Algorithm}
The complete flow of the achievable scheme is given in Algorithm~\ref{alg1}. First, $E_1$ creates the sets of indices $\mathcal{J}_r, r\in[L]$ and the vector $\bm{\alpha}$ of length $L$. Next, it executes CarPSI($X_{\mathcal{J}_1},N_2$) to find $M_{i_1}$ for the best function value $f_{i_1}$. Based on $M_{i_1}$ and $\alpha_1$, there are 3 cases:
\begin{itemize}
    \item [i)] If $M_{i_1}=0$, $E_2$ does not possess any element in $\mathcal{P}_2$ in common with $E_1$ that yields the same function value $f_{i_1}$. $E_1$ proceeds to the next best function value $f_{i_2}$ and finds $M_{i_2}$ by executing CarPSI($X_{\mathcal{J}_2},N_2$). 
    \item [ii)] If $M_{i_1}=\alpha_1$, then $f_{i_1}$ is a feasible function value with $\mathcal{P}_{alph}(\mathcal{J}_1)$ as the solution set. 
    \item [iii)] If $1\leq M_{i_1}\leq \alpha_1 -1$, $E_1$ proceeds to find the solution set which consists of $M_{i_1}$ out of $\alpha_1$ indices in $\mathcal{J}_1$ by implementing FindPSI($\mathcal{J}_1,M_{i_1},N_2$).
\end{itemize} 

In general,  CarPSI is executed according to i) till the returned cardinality $M_{i_r}\geq 1$, $r\in [L]$. If $\alpha_r=M_{i_r}$, $E_1$ directly sets $\mathcal{P}^*=\mathcal{P}_{alph}(\mathcal{J}_r)$. Otherwise, $E_1$ executes FindPSI. If the search continues and $M_{i_r}=0$ for every $r\leq L-1$, the solution is contained in $\mathcal{P}_{alph}(\mathcal{J}_L)$, since we assumed that $|\mathcal{P}_1\cap \mathcal{P}_2| \geq 1$. 

\begin{remark}
    One can view the aforementioned algorithm as a successive implementation of a ThPSI scheme with threshold $t=1$, on disjoint subsets of $\mathcal{P}_1$ corresponding to $\mathcal{J}_r$, $r\in [L]$ till the threshold is met on the intersection cardinality. This is done by computing $M=X_1^TX_2$ with CarPSI($X_1,N_2$) on $\mathbb{F}_q$ having $q>\min(P_1,P_2)$, prime, and FindPSI($X_1,M,N_2)$ if $t\leq M<P_1$.
\end{remark}

\subsection{Proof of Privacy} 
Each individual database of $E_2$ observes an independent, uniformly random vector over $\mathbb{F}_q^K$, in all rounds of CarPSI and FindPSI. This preserves $E_1$'s privacy \eqref{eq:leakage_non-leader}. 

For $E_2$'s privacy \eqref{eq:leakage_leader}, we guarantee that $E_1$ learns no more than $\mathcal{I}_{nom,2}$. From the answers and queries, $E_1$ learns that the elements in $\mathcal{P}_1 $ corresponding to indices $\cup_{r=1}^{R-1}\mathcal{J}_r$ are all absent in $E_2$'s constraint set, $\mathcal{P}_2$,  i.e., $X_2$ is 0 at those indices, and the value of $X_2$ at indices $\mathcal{J}_R$. Thus, $E_1$ learns $X_2$ at $\cup_{r=1}^{R}\mathcal{J}_r$ which is a subset of the support of $X_1$ and
\begin{align}
    I(\mathcal{P}_2;Q_{1:N_2},A_{1:N_2},\mathcal{P}_1,f) 
    &= H(\mathcal{P}_2) -   H(\mathcal{P}_2|Q_{1:N_2},A_{1:N_2},\mathcal{P}_1,f) \\
    &= H(X_2)-H\big(X_2([K] \setminus \cup_{r=1}^{R}\mathcal{J}_r)\mid X_2(\cup_{r=1}^{R}\mathcal{J}_r)\big)\\
    &=H\big(X_2(\cup_{r=1}^{R}\mathcal{J}_r)\big), \label{infoleak}
\end{align}
where we used the fact that $X_i$ is a sufficient statistic for $\mathcal{P}_i$.  

\subsection{Communication Cost}
$R$ is the least value of $r\in[L]$ in Algorithm~\ref{alg1} at which $M_{i_r}\geq 1$.  The download cost for the sequential CarPSI is equal to $D_{PSI}(R,N_2)$ which is $\big\lceil\frac{RN_2}{N_2-1}\big\rceil$. This is the download cost if $M_{i_R}=\alpha_R$. 

Otherwise, let  $R=k(N_2-1)+l$, for some $k\in\mathbb{N}\cup \{0\}$. If $l=0$, the download cost in FindPSI is $\big\lceil\frac{(\alpha_R-1)N_2}{N_2-1}\big\rceil$ and that from CarPSI is $\frac{RN_2}{N_2-1}$. Their sum gives \eqref{dlcost1}. 
With $l\in[N_2-2]$, after CarPSI($X_{\mathcal{J}_R}, N_2$), $l+1$ out of $N_2$ databases  are used. The remaining $N_2-(l+1)$ databases continue to send answers for FindPSI using the same random vector for query and common randomness symbols, respectively, resulting in the partial download cost of
\begin{align}
    \bigg\lceil{\frac{RN_2}{N_2-1}}\bigg\rceil+N_2-l-1= (k+1)N_2. \label{dpart}
\end{align}
This is followed by downloads for $\alpha_R-1-(N_2-l-1)$ indices which costs $D_{PSI}(\alpha_R-N_2+l,N_2)$, which when added to \eqref{dpart}, yields
\begin{align}
    D &= (k+1)N_2 + \bigg\lceil\frac{(\alpha_R-N_2+l)N_2}{N_2-1}\bigg\rceil\\
      &=\bigg\lceil\frac{(\alpha_R-1)N_2+(k(N_2-1)+l)N_2}{N_2-1}\bigg\rceil\\
      &= \bigg\lceil\frac{(\alpha_R-1+R)N_2}{N_2-1}\bigg\rceil.\label{lst}
\end{align}
There are $D$ answers downloaded, each of which is a response to a query of $K$ symbols, hence, the upload cost $U=KD$ and the communication cost  $C =(K+1)D$.

\subsection{Common Randomness Cost}
In the scheme, $E_1$ requires $\lceil{\frac{R+\alpha_R-1}{N_2-1}}\rceil$ random vectors for generating queries. This corresponds to equal amount of common randomness symbols $\mathcal{S}$ shared between the databases of $E_2$. However, since $R$ and $\alpha_R$ are not known beforehand, the databases of $E_2$ share a set of common randomness symbols $\mathcal{S}=\{S_1,S_2,\ldots, S_{m}\}$ where $m=\lceil{\frac{P_1}{N_2-1}}\rceil$, independent of $f$, $\mathcal{P}_1$ and $\mathcal{P}_2$. The databases do not collude hereafter. 

\subsection{Comparison with the Naive Scheme}
We now compare our scheme with a scheme where $E_1$ first finds the joint feasible set $\mathcal{P}=\mathcal{P}_1\cap \mathcal{P}_2$ privately, following the existing PSI protocol in \cite{zhushengPSI}, and then evaluates $f$ at the values in $\mathcal{P}$ only, selects the best ones among them to obtain $\mathcal{P}^*$. Finally, $E_1$ conveys this directly to $E_2$. We refer to this as the \emph{naive scheme}. 

Here, the entire $\mathcal{P}$ is leaked to $E_1$. Thus, the values of $X_2$ at indices in $\mathcal{J}_r$ $\forall r\in [L]$ are revealed to $E_1$ which is the support set of $X_1$. Thus,
\begin{align}
    I(\mathcal{P};\mathcal{P}_2|\mathcal{P}_1,f)= H(X_2(\cup_{r=1}^L \mathcal{J}_r))\geq H(X_2(\cup_{r=1}^R \mathcal{J}_r))=\mathcal{I}_{nom,2}.
\end{align} 
Equality holds only if $R=L$, which is when $M_{i_r}=0$ for all $r\in [L-1]$.

Since the upload cost is $K$ times the download cost in both schemes, it suffices to compare the communication efficiency in terms of the download cost. On one hand, the naive scheme requires $\lceil{\frac{P_1N_2}{N_2-1}}\rceil$ downloaded symbols of $\mathbb{F}_q$. For Example \ref{motivex}, the download cost in the naive approach is $\lceil{\frac{5N_2}{N_2-1}}\rceil$, irrespective of $f$. 

\begin{remark}\label{schemeNstar}
    Alternative to our proposed scheme, applying the scheme in \cite{zhushengPSI} sequentially over the elements in $\mathcal{P}_1$ corresponding to indices $\mathcal{J}_1$, $\mathcal{J}_2$ till $\mathcal{J}_R$ also reveals $\mathcal{I}_{nom,2}$ to $E_1$. The download cost incurred for this scheme is $\lceil\frac{(\sum_{r=1}^{R}\alpha_r)N_2}{N_2-1}\rceil$. Clearly, it is less communication-efficient due to higher download cost compared to $D$, except when $\alpha_r=1$ for all $r\in [R-1]$, in which case, they are equal.
\end{remark}

\begin{remark}
    Under uniform random mapping of $f$, with a fixed realization of $\mathcal{P}_1$ and $\mathcal{P}_2$, the probability $P_{eq}$ that the download costs $D$ and $D_{PSI}$ are equal, is small. To support this, we provide the following calculation, and a numerical result.

    With a fixed realization of $\mathcal{P}_1$ and $\mathcal{P}_2$, the random variable $R\in [L]$ and vector $\bm{\alpha}$ depend on the realization of $f$. The maximum value of $L$ (hence $R$) is $R_{\max}=\min(\tau,P_1-M+1)$. Further, for any $R\in [L]$, $M_{i_R}\leq\alpha_R$. If $M_{i_R}=\alpha_R$, we drop the FindPSI phase, and the download cost is $\lceil \frac{RN_2}{N_2-1}\rceil$. For this to equal $D_{PSI}$, we require $R=P_1$, which implies $M=1$. Hence, $R=L$ and our algorithm skips the last CarPSI step as well, in which case the actual download cost is $\lceil \frac{(L-1)N_2}{N_2-1}\rceil$. As a result, $D_{PSI}>D$ when $M_{i_R}=\alpha_R$. 
    
    Therefore, $M_{i_R}$ must be strictly less than $\alpha_R$ so that FindPSI is executed. Thus, $D=D_{PSI}$ if $\{R+\alpha_R=P_1+1\}$ and $\{M_{i_R}<\alpha_R\}$ are simultaneously satisfied. Note that, for such events $\alpha_r=1$ for all $r<R$. Further, $M_r=0$ for all $r<R$, hence the function value of every element in $\mathcal{P}=\mathcal{P}_1\cap \mathcal{P}_2$ is $f_{i_R}$, i.e., optimum. With $M \geq 1$, the corresponding probability $P_{eq}$ is,
    \begin{align}\label{Mralphar} 
        P_{eq}&=\mathbb{P}(R+\alpha_R=P_1+1, M_{i_R}<\alpha_R)\\
        &=\sum_{r=1}^{R_{\max}}\mathbb{P}(R=r,\alpha_r=P_1+1-r, M_{i_R}<P_1+1-r) 
        \label{Malphar}\\
        &=\sum_{r=1}^{R_{\max}}\mathbb{P}(R=r,\alpha_r=P_1+1-r)\mathbb{P}(M<P_1+1-r)
        \\
        &=\sum_{r=1}^{R_{\max}}P_{eq}(r)\times\mathds{1}_{\{M<P_1-r+1\}}, \label{indic}
    \end{align}
    where \eqref{Malphar} follows from \eqref{Mralphar} using $\mathbb{P}(M_{i_R}<P_1+1-r|R=r,\alpha_r=P_1+1-r)=\mathbb{P}(M<P_1+1-r)$ since $\mathcal{P}=\mathcal{P}^*$. Further, since $M$ is determined by $\mathcal{P}_1$ and $\mathcal{P}_2$, $\mathbb{P}(M<P_1+1-r)$ is an indicator as in \eqref{indic}. If $r=1$, we have
    \begin{align}
        P_{eq}(1)=\tau\bigg(\frac{1}{\tau}\bigg)^{P_1},
    \end{align}
    since $\tau$ out of $\tau^{P_1}$ function realizations assume equal function value over $\mathcal{P}_1$. For $1<r\leq R_{\max}$, 
    \begin{align}\label{peqr}
        P_{eq}(r)=&\binom{P_1-M}{r-1}(r-1)!\bigg[\sum_{j=r-1}^{\tau-1}\binom{j-1}{r-2}\times(\tau-j)\bigg]\times\bigg(\frac{1}{\tau}\bigg)^{P_1}.  
    \end{align}
    
    \begin{figure}[t]
        \centering
        \includegraphics[width=0.6\textwidth]{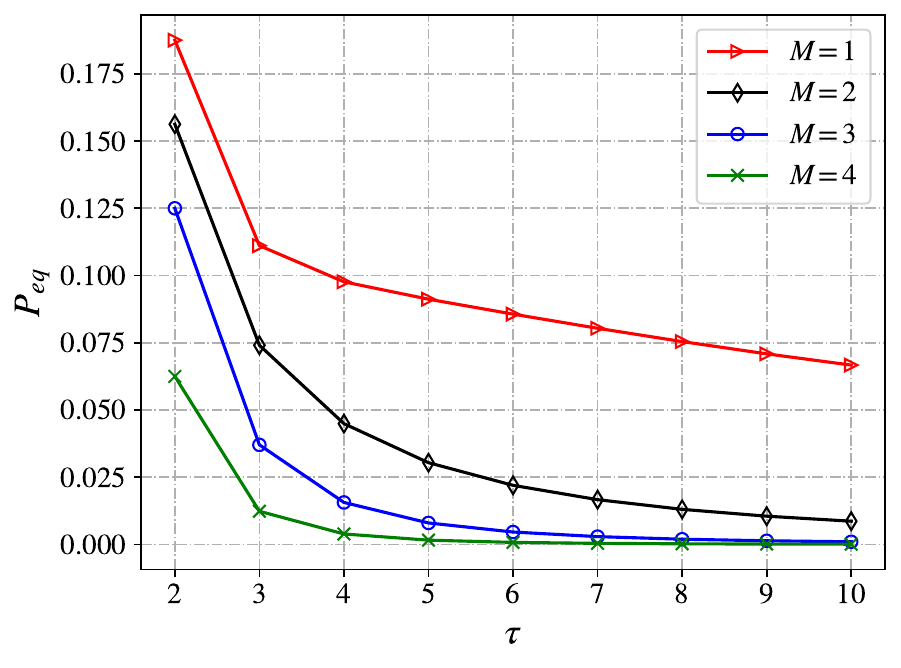}
        \caption{$P_{eq}$ against $\tau$ for different values of $M$ with $N=2$.}
        \label{probTM}
    \end{figure}
    
    The term preceding the summation in \eqref{peqr} follows by sampling $f_{i_1},f_{i_2},\ldots,f_{i_{r-1}}$ with ordering from the $P_1-M$ elements in $\mathcal{P}_1$ absent in $\mathcal{P}$. Fixing $f_{i_{r-1}}=j$, we choose $f_{i_1},\ldots,f_{i_{r-2}}$ from $1$ to $j-1$, as better function values preceding $f_{i_{r-1}}$, in $\binom{j-1}{r-2}$ ways. The remaining $\alpha_R=P_1-r+1$ elements take the same function value $f_{i_R}$, which has $\tau-j$ choices in the set $\{j+1,j+2,\ldots,\tau\}$ and we sum over all  $j\in\{r-1,\ldots,\tau-1\}$. This expression, multiplied by the uniform probability of function realization, yields \eqref{peqr}.

    To show an example that \eqref{indic} is small, in Fig.~\ref{probTM}, we plot the probability $P_{eq}$ with respect to the size of function range $\tau$ for various values of $M$. We follow the example of movie scores in Section~\ref{motivex} and fix $P_1=5$ while the scores are allotted from the set $\mathcal{T}=[\tau]$ with $\tau\in\{2,3,\ldots,10\}$. For a fixed $\tau$, increasing $M$ from $1$ to $4$ decreases the value of $P_{eq}$. 
\end{remark}

\section{The Case of $N>2$ Entities: Ring Setup}\label{achievable>2_ring}
The underlying idea of the achievable scheme, similar to the case of $N=2$, is to apply ThPSI with $t=1$ on each set of equi-cost elements in $\mathcal{P}_{alph}$ sequentially, till the threshold condition of non-empty intersection is satisfied. We present a PSI scheme under the ring setup, and show that the communication cost of our DOFSP scheme is less than or equal to that of the PSI scheme. Through numerical results, we show that the probability that the communication costs of the two schemes are equal, is low. 

We begin the description our scheme with an example.

\begin{example}\label{example_dofsp_ring}
    Continuing with the movie example, let $\mathcal{P}_{alph}=\{A,B,C,D,E,F,G,H,I,J\}$ be the set of movies and $f$ be their scores assigned as follows:
    \begin{align}
       f(x)=
       \begin{cases}
          1, & x\in \{A\},\\
          2, & x\in \{B,E\},\\
          3, & x\in \{F,G,H\},\\
          4, & x\in \{I,J\},\\
          5, & x\in \{C,D\}.
       \end{cases}
    \end{align}
    Then, with $N=4$ entities, let the individual feasible sets be as follows
    \begin{align}
        \mathcal{P}_1&=\{A,C,D,G,J\}
        &\implies X_1=[1, 0, 1, 1, 0, 0, 1, 0, 0, 1]^T , \hat{X}_1=[1, 1, 0, 1, 0, 1, 0, 0, 0, 1]^T \nonumber\\
        \mathcal{P}_2&=\{B,E,G,H,I,J\}&\implies X_2=[0, 1, 1, 0, 1, 0, 1, 1, 1, 1]^T, \hat{X}_2=[0,0,1,1,0,1,1,1,1,0]^T \nonumber \\
        \mathcal{P}_3&=\{A,B,C,F,G,I\}&\implies X_3=[1, 1, 1, 0, 0, 1, 1, 0, 1, 0]^T , \hat{X}_3=[1,0,1,0,1,1,0,1,0,1]^T \nonumber \\
        \mathcal{P}_4&=\{C,D,G,H,I\}&\implies X_4=[0, 0, 1, 1, 0, 0, 1, 1, 1, 0]^T , \hat{X}_4=[1,1,1,0,0,1,1,0,0,0]^T 
    \end{align}
    where $\hat{X}_i$ is $X_i$ after permuting the indices as 
    \begin{align}
       \hat{X}_i(1,2,3,4,5,6,7,8,9,10) = X_i(3,4,9,10,6,7,8,2,5,1). 
    \end{align} Let $N_i=2 $ for each $E_i$, implying $\mathcal{N}_{eff}=\{1,2\}$. Let $q=5$, and $E_i, i\in \{2,3,4\}$ share uniformly and independently generated randomness vectors $S^{(1)}_i\in \mathbb{F}_q^2, S^{(2)}_i\in \mathbb{F}_q^2, S^{(3)}_i\in \mathbb{F}_q^3, S^{(4)}_i \in \mathbb{F}_q^2$ and $S^{(5)}_i\in \mathbb{F}_q$ among each database pair. First, the entities check for the presence of $\{C,D\}$ (elements that map to $f_1$) in $\mathcal{P}$. Towards this, $E_1$ generates $h_1$ uniformly at random from $\mathbb{F}_q^2$ to send the queries
    \begin{align}
        Q^{(1)}_{1,1}&=h_1, \quad Q^{(1)}_{1,2}=h_1+\hat{X}_1(1,2), \\
        Q^{(1)}_{2,j}&=Q^{(1)}_{1,j}\circ \hat{X}_2(1,2) +S^{(1)}_2, \quad
        Q^{(1)}_{3,j}=Q^{(1)}_{2,j}\circ \hat{X}_3(1,2) +S^{(1)}_3, \quad j\in \{1,2\} \\
        A^{(1)}_1 &= {Q^{(1)}_{3,1}}^T\hat{X}_4(1,2)+S^{(1)}_4(2), \quad {Q^{(1)}_{3,2}}^T\hat{X}_4(1,2)+S^{(1)}_4(2),
    \end{align}
    where the superscript indicates the round index, and the symbol $\circ$ indicates elementwise multiplication. Next, $E_1$ evaluates 
    \begin{align}
        A^{(1)}_2-A^{(1)}_1=&{(((h_1+\hat{X}_1(1,2))\circ \hat{X}_2(1,2)+S^{(1)}_2)\circ\hat{X}_3(1,2)+S^{(1)}_3)}^T \hat{X}_4(1,2)+S^{(1)}_4(2)\nonumber \\
        &-{((h_1\circ \hat{X}_2(1,2)+S^{(1)}_2)\circ\hat{X}_3(1,2)+S^{(1)}_3)} ^T \hat{X}_4(1,2)+S^{(1)}_4(2)\\
        =&(\hat{X}_1(1,2)\circ \hat{X}_2(1,2) \circ \hat{X}_3(1,2))^T\hat{X}_4(1,2)=0.
    \end{align}
    Since the intersection is empty, no elements mapped to $5$ are present in $\mathcal{P}$. Now, $E_1$ proceeds to the next best value $4$ and generates a new random vector $h_2\in \mathbb{F}_q^2$ and similarly evaluates $A^{(2)}_2-A^{(2)}_1=(\hat{X}_1(3,4)\circ \hat{X}_2(3,4) \circ \hat{X}_3(3,4))^T\hat{X}_4(3,4)$ which is also $0$. Next, $E_1$ generates  $h_3\in \mathbb{F}_q^3$ and similarly evaluates $ A^{(3)}_2-A^{(3)}_1=(\hat{X}_1(5,6,7)\circ \hat{X}_2(5,6,7) \circ \hat{X}_3(5,6,7))^T\hat{X}_4(5,6,7))=1$, which is the first non-zero cardinality. Now, $E_1$ sends two elements from $\mathbb{F}_q$, one each to the databases of $E_4$, which signals them to send the following answers to $E_1$,
    \begin{align}
        A_j'=(Q^{(3)}_1\circ \hat{X}_4)(5,6)+S^{(3)}_4(1,2), \quad j\in \{1,2\}.
    \end{align}
    Using these, $E_1$ calculates $A_2'-A_1'=\hat{X}_1(5,6)\circ \hat{X}_2(5,6) \circ \hat{X}_3(5,6,)\circ \hat{X}_4(5,6)=[0,1]^T$. This gives $\mathcal{P}^*=\{G\}$, since the cardinality is $1$. The communication protocol stops after $3$ rounds where each of the first two rounds required $2\cdot3\cdot2+2=14$ symbols and the last round required $2\cdot3\cdot3+2+2+4=26$ symbols, hence $C_{ring}=14+14+26=54$. 
\end{example}

\subsection{Scheme Preliminaries}
First, we extend the two-entity ($N=2$) functionalities CarPSI and FindPSI from Section~\ref{achievable2} to the ring setup, to develop new primitives CarPSI-ring and FindPSI-ring.\footnote{Similar to the $N=2$ case, these primitives can be executed to attain a ThPSI scheme. After CarPSI-ring, if the resulting cardinality $M$ is $t$ or greater, $E_1$ executes FindPSI-ring, unless $M=P_1$, which is the trivial case that $\mathcal{P}=\mathcal{P}_1$.} In CarPSI-ring, $E_1$ privately obtains the cardinality $M$ given by,
\begin{align}\label{PSIcar}
    M=\bm{1}^T(X_1\circ X_2 \ldots \circ X_N)=(X_1\circ X_2 \ldots \circ X_{N-1})^T X_N,
\end{align}
without learning the actual intersection $\mathcal{P}$, 
where $\mathbf{1}$ is the $K\times 1$ all ones vector and $\circ$ denotes the elementwise product operation. FindPSI-ring is executed after this, if the cardinality exceeds a certain value. We assume that each entity has at least two replicated non-colluding databases storing the incidence vectors of the feasible sets. Before the start of CarPSI-ring, the entities agree on an operating field $\mathbb{F}_q$ where $q>\min_{i\in[N]} P_i$ and prime. We let the shared common randomness $\mathcal{S}_i$ be a random vector $S_i\in \mathbb{F}_q^K$ whose entries $S_i(k)$ are picked uniformly at random and independently from $\mathbb{F}_q$ for each $i\in [2:N]$ and $ k\in [K]$. 

\subsubsection{CarPSI-ring($X_1$)}\label{carpsi_ring}
First, $E_1$, using a private random vector $h$ drawn uniformly from $\mathbb{F}_q^K$ generates two queries $Q_{1,1}=h_1$ and $Q_{1,2}=h_1+X_1=Q_{1,1}+X_1$  and sends them to databases $1$ and $2$ of $E_2$, respectively. On receiving $Q_{1,j}$ database $j=1,2$ computes its elementwise product with $X_2$, and adds the randomness vector $S_2$ to it. Then, database $j\in \{1,2\}$ sends the computed value as the query $Q_{2,j} = Q_{1,j}\circ X_2+S_2$ to database $j$ of $E_3$, and so on. In general, for each $i\in [N-2]$, database $j$ of $E_i$ uses its incoming query to send the following to database $j$ of $E_{i+1}$, 
\begin{align}
    Q_{i+1,j}=&Q_{i,j}\circ X_i + S_{i}\\
    =&(Q_{i-1,j}\circ X_{i-1}+S_{i-1})\circ X_i + S_{i}\\
    &\vdots \nonumber\\
    =&((Q_{1,j}\circ X_2 +S_2)\circ X_3+S_3)\circ \ldots \circ X_i+ S_{i}.
\end{align}
Finally, the databases of $E_N$ use their incoming queries $Q_{N-1,j}$ to compute and send the following answers to $E_1$,
\begin{align}
    A_1&= {Q_{N-1,1}}^T X_N + S_{N}(K), \\
    A_2&= {Q_{N-1,2}}^T X_N +S_{N}(K),
\end{align}
where $S_N(K)$ is the $K$th coordinate of $S_N$. From the received answers, $E_1$ computes $A_2-A_1$. 

To prove that the leader obtains $M$, we first prove the fact that, for all $i\in [N-1]$,
\begin{align}\label{eq:fact}
    Q_{i,2}=Q_{i,1}+(X_1\circ\ldots \circ X_i).
\end{align}
We show this by induction on $i$. For $i=1$, it is trivially true. Assume that this is true for $i=k$ for some $k\in [2:N-2]$. Then, we have,
\begin{align}
    Q_{k+1,2} & = Q_{k,2} \circ X_{k+1} + S_{k+1}\\
    & = \left(Q_{k,1} + (X_1\circ \ldots \circ X_k)\right) X_{k+1} + S_{k+1} \label{ind_hypothesis}\\
    & = Q_{k,1}\circ X_{k+1} + (X_1\circ \ldots \circ X_{k+1}) + S_{k+1} \\
    & = Q_{k+1,1} + (X_1\circ \ldots \circ X_{k+1}),
\end{align}
as required. Therefore, 
\begin{align}
    A_2-A_1 &=(Q_{N-1,2}-Q_{N-1,1})^T X_N+S_N(K)-S_N(K)\\
    &=(X_1\circ X_2\circ\ldots \circ X_{N-1})^T X_N. \label{proof_correct}
\end{align}

To prove privacy, note that the queries sent by $E_1$ to $E_2$ protect the privacy of $X_1$ (hence, of $\mathcal{P}_1$) since,
\begin{align}
    I(Q_{1,1};X_1|{S}_2, X_2)&=I(h_1;X_1)=0, \label{E1priv1}\\
    I(Q_{1,2};X_1|{S}_2, X_2)&=I(h_1+X_1;X_1)=0,\label{E1priv2}
\end{align}
where \eqref{E1priv1} is true because $h_1$, ${S}_2$ and $X_2$ are independent of $X_1$, and \eqref{E1priv2} uses Shannon's one-time padding \cite{shannonotp}.
For $i\in [2:N-1]$, the privacy of $E_1$ through $E_i$ is preserved against an individual database of $E_{i+1}$. We show this by induction on $i$. Let $i=2$, for database $1$,
\begin{align}
   I(Q_{2,1};X_1,X_2|{S}_3, X_3)&=I(h_1\circ X_2 +S_2;X_1,X_2) \label{eq:omit_S3_X3} \\
   &=I(h_1\circ X_2 +S_2;X_1)+I(h_1\circ X_2 +S_2;X_2|X_1) \label{E12priv1} \\
   &=0,
\end{align}
where $S_3,X_3$ are dropped in \eqref{eq:omit_S3_X3} since they are independent of $Q_{2,1}, X_1$ and $X_2$. In \eqref{E12priv1}, the first term is $0$ because $h_1, X_2$ and $S_2$ are independent of $X_1$ and the second term is $0$ by one-time padding with $S_2$. Similarly, for database $2$, 
\begin{align}
   I(Q_{2,2};X_1,X_2|{S}_3, X_3)&=I((h_1+X_1)\circ X_2 +S_2;X_1,X_2) \\
   &=I((h_1+X_1)\circ X_2 +S_2;X_1)+I((h_1+X_1)\circ X_2 +S_2;X_2|X_1)\\
   &=0.
\end{align}
Let $I(Q_{i-1,j};X_1,\ldots,X_{i-1}|X_{i}, S_{i})=0$, $\forall j=1,2$, then, 
\begin{align}
    I(Q_{i,j};X_1,\ldots,X_i|X_{i+1}, S_{i+1})=&  I(Q_{i-1,j}\circ X_{i}+S_{i};X_1,\ldots,X_{i-1}) \nonumber \\
    &+ I(Q_{i-1,j}\circ X_{i}+S_{i};X_i|X_1,\ldots,X_{i-1}) \label{E1ipriv}\\
    =&0, 
\end{align}
where the first term in \eqref{E1ipriv} is 0 by the induction hypothesis and the fact that the conditioning terms, being independent, can be dropped. The second term in \eqref{E1ipriv} is 0 by one-time padding. The answers $A_1$ and $A_2$ are one-time padded with the $K$th entry of $S_N$, hence, they appear to be chosen uniformly at random from $\mathbb{F}_q$. The evaluation $A_2-A_1$ reveals only $M$ to $E_1$ and maintains the privacy of $\mathcal{P}$. 
The communication cost for CarPSI-ring is $2(N-1)K+2$ symbols.
\begin{remark}
    When $N=2$, CarPSI-ring reduces to CarPSI($X_1, N_2$) as described in Section \ref{sec:CarPSI_N=2} with $M=X_1^T X_2$.
\end{remark}

\subsubsection{FindPSI-ring($M$)}
Here, $E_1$ sends back two random symbols of $\mathbb{F}_q$, one to each of the responding databases (say 1 and 2) of $E_N$ to signal them for answers. This time, the databases of $E_N$ reuse the received queries from CarPSI-ring, and respond with,
\begin{align}
    A_1'&=(Q_{N-1,1}\circ X_N)(1:K-1) +S_N(1:K-1), \\
    A_2'&=(Q_{N-1,2}\circ X_N)(1:K-1) +S_N(1:K-1).
\end{align}
Thus, the answers are ($K-1$)-length vectors since the last entries of $Q_{N-1,j}$ and $X_N$ are not included. We denote the incidence vector of $\mathcal{P}$ by $X_{\cap}$, i.e., $X_{\cap}=X_1\circ X_2 \circ \ldots \circ X_N$. $E_1$ directly obtains $X_{\cap}(1:K-1)$ from $A_2'-A_1'$ due to \eqref{eq:fact}. This completes the retrieval of $X_{\cap}$, and $E_1$ learns the intersection $\mathcal{P}$, since $M=\bm{1}^T X_{\cap}$ is already known to it.

To show privacy, $A_1'$ and $A_2'$ individually hide $X_N$ because each of them is one-time padded with $S_N(1:K-1)$, which is independent of $S_N(K)$. After computing $A_2'-A_1'$, $E_1$ only learns the function $X_2 \circ \ldots \circ X_N$ and nothing beyond on $X_2,\ldots, X_N$ because of \eqref{E1priv2}. Further, the databases of $E_N$ receive random symbols from $E_1$, which reveal no more information than the threshold condition being satisfied. The communication cost is $2+2(K-1)=2K$, where the first term is contributed by the symbols sent by $E_1$ and the second term arises from the answers of $E_N$.

With these primitives at hand, we proceed with the steps of our DOFSP scheme.
\subsection{Scheme Description}
Let $f_R$ be the optimum function value for the given objective. Then, to retrieve $\mathcal{P}^*$, $E_1$ executes the CarPSI-ring primitive over partitions of $X_i$ till the $R$th round and one FindPSI-ring primitive. We denote the query $Q_{i,j}$ as the collection of all queries $Q^{(r)}_{i,j}, \forall r\in [R]$ sent over $R$ rounds. Similarly, $A_j=(\cup_{r=1}^R A^{(r)}_j)\cup A'_j$. The operating field $\mathbb{F}_q$ is chosen such that $q$ is a prime greater than $\max_{r\in [T]}\mu_r$. This is to ensure correct evaluation of the set intersection cardinalities in the CarPSI-ring primitive.

\paragraph{Partitioning $X_i$ into Sub-Alphabets:} Each entity $E_i$ groups $X_i$ into $T$ partitions, where each partition comprises the equi-cost elements. For the objective function $f$, the entities use a fixed permutation $\Pi_f:[K]\to [K]$, $X_i$ is rearranged into $\hat{X}_i:=\Pi_f(X_i)$ for each $i\in [N]$. For each partition $r\in [T]$, let $\mathcal{I}_r:=(k_1,k_2,\ldots,k_{\mu_r})$, with $k_1<k_2<\ldots<k_{\mu_r}$ denote an ordered set such that $f(\mathcal{P}_{alph}(k_1))=f(\mathcal{P}_{alph}(k_2))=\ldots=f(\mathcal{P}_{alph}(k_{\mu_r}))=f_r$, which is the $r$th best function value. Then, 
\begin{align}
   \Pi_f(k_j) = \sum_{v=1}^{r-1}{\mu_v}+j \implies \hat{X}_i\Big(\sum_{v=1}^{r-1}{\mu_v}+j\Big)=X_i(k_j), \qquad \forall j \in [\mu_r].
\end{align}

Note that, the permutation $\Pi_f$ is known to all entities since it is derived directly from $f$. Let, for each $i\in [N]$, $X(\mathcal{I}_r)_{\mu_r\times 1}$ denote the column vector with entries at  $\mathcal{I}_r$, then
\begin{align} 
    \hat{X}_i=\begin{bmatrix}
    X_i(\mathcal{I}_1)^T & X_i(\mathcal{I}_2)^T \ldots & X_i(\mathcal{I}_T)^T
    \end{bmatrix}^T,
\end{align}
is stored across the databases in $\mathcal{N}_{eff}$ databases for each $E_i$. 

\paragraph{Assignment of Common Randomness:} The databases in $\mathcal{N}_{eff}$ of non-leader entities generate and share common randomness vectors as follows.

\begin{enumerate}
    \item If $N_{eff}=2$, databases $1$ and $2$ of $E_2$ through $E_N$ generate randomness vectors uniformly at random from $\mathbb{F}_q^{\mu_r}$ for each $r\in [T]$, where $r$ denotes the round of communication. The vectors are denoted by $S^{(r)}_i$, $i\in [2:N]$.
    \item Next, if $N_{eff}>2$, the databases $\{j, j+1\}\in\mathcal{N}_{eff}$ of $E_i$ are assigned the rounds $r\in \mathcal{R}_j$ in a cyclic fashion, modulo $N_{eff}$, corresponding to which they store $S_i^{(r)}\in \mathbb{F}_q^{\mu_{r_j}}$. Specifically, for each $i\in [2:N]$: 
        \begin{itemize}
        \item If $N_{eff}$ is even,
        \begin{align}\label{setrj1}
        \mathcal{R}_j=\left\{kN_{eff}/2+\big\lceil\frac{j}{2}\big\rceil,\ k\in \left[0:\bigg\lfloor\frac{2(T-\lceil\frac{j}{2}\rceil)}{N_{eff}}\bigg\rfloor\right]\right\}.
        \end{align}
        \item If $N_{eff}$ is odd, 
        \begin{align}\label{setrj2}
            \mathcal{R}_j=
            \begin{cases}
                 \left\{\big\lfloor kN_{eff}/2 \big\rfloor+\frac{j+1}{2},\ k\in \left[0:\big\lceil\frac{2T-j-1}{N_{eff}}\big\rceil\right]\right\},    & j \hbox{ is odd},\\
                 \left\{\big\lceil kN_{eff}/2\big\rceil+\frac{j}{2},\ k\in \left[0:\big\lfloor\frac{2T-j}{N_{eff}}\big\rfloor\right]\right\}, & j \hbox{ is even}.
            \end{cases}
        \end{align}
        \end{itemize}
    For example, if $T=5$ and $N_{eff} = 4$, $\mathcal{R}_1 = \mathcal{R}_2 = \{1,3,5\}$ and $\mathcal{R}_3=\mathcal{R}_4 = \{2,4\}$. On the other hand, if $N_{eff} = 3$, $\mathcal{R}_1 = \{1,2,4,5\}$, $\mathcal{R}_2 = \{1,3,4\}$ and $\mathcal{R}_3 = \{2,3,5\}$.
\end{enumerate}

\paragraph{Query and Answer Generation:} If $N_{eff}>2$, database $j$ of $E_2$ through $E_N$ send queries and answers for rounds in $\mathcal{R}_j$ as given by \eqref{setrj1} and \eqref{setrj2}. If $N_{eff}=2$, then each database participates in all the rounds. 

Next, we provide a description of our scheme for $N_{eff}=2$ and that for $N_{eff}>2$ can be similarly derived, taking care that, a distinct random vector is used, each time the same database is queried. The quantity $N_{eff}$ does not affect the overall communication, computation or randomness costs, and only affects the load on individual databases. That is, a higher value of $N_{eff}$ reduces the load of computation, since each database receives queries for fewer number of rounds.

In the first round, $E_1$ executes CarPSI-ring on $\hat{X}_1([\mu_1])$. That is, it generates a vector $h_1$ uniformly at random from $\mathbb{F}_q^{\mu_1}$ and sends $Q^{(1)}_{1,1}=h_1$ and $Q^{(1)}_{1,2}=h_1+\hat{X}_1([\mu_1])$ to databases $\{1,2\}$ of $E_2$. Next, database $j$ of $E_2$ sends $Q^{(1)}_{2,j}=Q^{(1)}_{1,j}\circ \hat{X}_2([\mu_1])$ to database $j$ of $E_3$. This continues for $i\in [2:N-1]$ following which $E_N$ sends answers to $E_1$ as follows
\begin{align}
    A^{(1)}_j={Q^{(1)T}_{N-1,1}} \hat{X}_N([\mu_1])+S^{(1)}_{N}(\mu_1), \quad j=1,2.
\end{align} 
Using these answers, $E_1$ evaluates $\hat{X}_{\cap}(\mu_1)=A^{(1)}_2-A^{(1)}_1$ to retrieve $M_1$, i.e., the cardinality of elements in $\mathcal{P}$ that take the value $f_1$. If this is $0$, $E_1$ executes the next round of CarPSI-ring.

In round $r\geq 2$, $E_1$ generates a fresh random vector $h_r\in \mathbb{F}_q^{\mu_r}$ to send the queries $Q^{(r)}_{1,1}=h_r$ and $Q^{(r)}_{1,1}=h_r+\hat{X}_1\big(\big[\sum_{v=1}^{r-1}\mu_v+1: \sum_{v=1}^{r-1}\mu_v +\mu_r\big]\big)$. The same set of operations are followed by the subsequent entities and, for $i\in [N-1]$, $j\in \mathcal{N}_{eff}$,
\begin{align}
   Q^{(r)}_{i+1,j}=Q^{(r)}_{i,j}\circ\hat{X_i}\big(\big[\sum_{v=1}^{r-1}\mu_v+1:\sum_{v=1}^{r-1}\mu_v +\mu_r\big]\big)+S^{(r)}_i(\mu_r).
\end{align} 
Further, the answer returned by database $j\in \mathcal{N}_{eff}$ of $E_N$ is,
\begin{align}
    A_j^{(r)}={Q^{(r)T}_{N-1,j}} \hat{X}_N\big(\big[\sum_{v=1}^{r-1}\mu_v+1:\sum_{v=1}^{r-1}\mu_v +\mu_r\big]\big) + S^{(r)}_N(\mu_r).
\end{align}
Let $r=R$ denote the round for which $M_r\geq 1$ for the first time. Now, $E_1$ executes FindPSI-ring($M_R$), receiving the answers,
\begin{align}
   A_j'= (Q^{(R)}_{N-1,j}\circ \hat{X}_N)([\sum_{v=1}^{R-1}\mu_v+1:\sum_{v=1}^{R-1}\mu_v+\mu_R-1\big])+S^{(R)}_N([\mu_{R}-1]), \quad j=1,2.
\end{align}
Next, $E_1$ evaluates $A_2'-A_1'$ to obtain $\hat{X}_1\circ \ldots \circ \hat{X}_N$ at indices $[\sum_{v=1}^{R-1}\mu_v+1:\sum_{v=1}^{R}\mu_v-1\big]$. From this and $M_R$, 
$\hat{X}_{\cap}(\mu_R)$ is retrieved. Let $\mathcal{J}^*$ denote the support of $\hat{X}_{\cap}(\mu_R)$, then $E_1$ evaluates $\mathcal{P}^*=\mathcal{P}_{alph}(\mathcal{J}^*)$. 

\begin{remark}\label{reducedcost1}
    If $\mu_R=1$ or $M_R=\mu_R$, then $\mathcal{P}^*=\mathcal{P}_{alph}(\mathcal{I}_R)$ is found directly without executing FindPSI-ring($M_R$).
\end{remark}

\begin{remark}\label{reducedcost2}
    If $R=T$ and $\mu_T>1$, then $E_1$ executes FindPSI-ring, without sending the two signaling symbols to the databases of $E_N$, because of the assumption of non-empty intersection, $\mathcal{P}$. 
\end{remark}

\paragraph{Proof of Privacy:} The privacy of the scheme follows from the individual privacy guarantees of the CarPSI-ring and FindPSI-ring primitives, performed on disjoint subsets of $\mathcal{P}_{alph}$. Moreover, each database receives queries of length $\mu_r$ for the rounds it is assigned, till $R$ is found. This is regardless of the fact whether the value $f_r$ is realized by any element in $\mathcal{P}_1$. Therefore, no information about the leader's function profile $\bm{\alpha}$ is leaked to any database. There is no leakage to the leader $E_1$, beyond $\mathcal{I}_{nom,1}$. After round $r<R$, $E_1$ only learns that $\hat{X}_{\cap}(\mu_r)$ is empty and gains no information about the individual incidence vectors $\hat{X}_2([\mu_{R-1}])$ through $\hat{X}_N([\mu_{R-1}])$.

\paragraph{Communication Cost:} The communication cost is contributed by $R$ rounds of CarPSI-ring, and FindPSI-ring at the $R$th round. First, we consider the case that $R<T$. Then, if $\mu_R>1$,
\begin{align}
    C_{ring}&=\sum_{r=1}^R\left(2(N-1)\mu_r + 2\right) + 2 + 2(\mu_R -1)\\
    &=2N\left(\sum_{r=1}^R\mu_r\right)+2\left(R-\sum_{r=1}^{R-1}\mu_r\right). \label{comm_cost_ring}
\end{align}
If $\mu_R=1$, the cost reduces to $C_{ring} = 2N\left(\sum_{r=1}^R\mu_r\right)+2\left(R-\sum_{r=1}^{R}\mu_r\right)$ by Remark~\ref{reducedcost1}. Now, consider the case that $R=T$. If $\mu_R>1$, following Remark~\ref{reducedcost2}, two fewer symbols are required, which gives
\begin{align}
    C_{ring} &=2N\left(\sum_{r=1}^R\mu_r\right)+2\left(R-\sum_{r=1}^{R-1}\mu_r\right)-2 \\
    &=2NK+2\left(R-1-\sum_{r=1}^{R-1}\mu_r\right),
\end{align}
since $\sum_{r=1}^T\mu_r=K$. If $\mu_R=1$, no communication takes place, further saving $2\mu_R$ symbols and $C_{ring} =2NK+2\left(R-1-\sum_{r=1}^{R}\mu_r\right)$. This completes the proof of Theorem \ref{thm2}. 

\subsection{Comparison with the Naive Scheme} 
First, we describe the PSI scheme for the ring setup, obtained by modifying the CarPSI-ring primitive.

\paragraph{PSI Scheme:}\label{PSIring}
 In this scheme, the query generation process for entities $E_1$ through $E_{N-1}$ matches that of CarPSI-ring($X_1$) in Section~\ref{carpsi_ring}, and differs in the answers returned by $E_N$. The databases of $E_N$ generate answers from $\mathbb{F}_q^K$ as,
\begin{align}
    A_j= Q_{N-1,j}\circ X_N + S_{N},\quad j=1,2.
\end{align}
The communication cost thus incurred is 
\begin{align}\label{cost_psi_ring}
    C_{PSI,ring}=2NK
\end{align} symbols. The communication cost $C_{ring}$ is upper bounded by $C_{PSI,ring}$, with $P_{eq}$ denoting the probability that they are equal. 

To evaluate $P_{eq}$, we study the conditions under which the communication costs are equal. It suffices to consider that $\mu_R>1$, since $C_{ring}$ is strictly lower if $\mu_R=1$. First, if $R<T$ then, $C_{ring}$ equals $2NK$ if and only if
\begin{align}
    2NK=2N\sum_{r=1}^R\mu_r+2\left(R-\sum_{r=1}^{R-1}\mu_r\right),
\end{align}
which implies
\begin{align} 
    K-\sum_{r=1}^R\mu_r= \frac{R-\sum_{r=1}^{R-1}\mu_r}{N},
\end{align}
which gives
\begin{align} \label{eq:contradiction_ring}
    \sum_{r=R+1}^T\mu_r = \frac{R-\sum_{r=1}^{R-1}\mu_r}{N}.
\end{align}
However, note that, in \eqref{eq:contradiction_ring}, the left-hand side is a positive integer, with value at least $T-R>1$ since $\mu_r\geq 1$, $r\in [T]$, whereas, the right hand side is less than or equal to $\frac{1}{N}$, since the numerator is at most 1. This leads to a contradiction since $N\geq 2$. Thus, equality of costs happen only if $R=T$.

For $R=T$, if $\mu_R=\mu_T>1$,
\begin{align}
    C_{ring}&=2NK-2\left(K-\mu_T-(T-1)\right)\\
    &=2NK+2\left((T-1)-\sum_{r=1}^{T-1}\mu_r\right)\\
    &\leq 2NK,
\end{align}
since $\mu_r\geq 1$ by definition. Clearly, equality holds iff $\mu_r=1, \forall r\in [T-1]$ and $\mu_T>1$. Given the feasible sets $\mathcal{P}_1,\ldots,\mathcal{P}_N$ and $\tau$, depending on the realization of $f$, $T$ varies from $1$ to $T_{\max}$ given by $T_{\max}=\min(\tau,K-\max(2,M)+1)$. Since $T$ is upper-bounded by $\tau$, its maximum value when $\tau \leq K-\max(2,M)+1$ is $\tau$. Otherwise, the number of required communication rounds is $K-M+1$, since $T=R$ if $M>1$. If $M=1$, the last communication round is skipped and $K-1$ rounds suffice to find $\mathcal{P}^*$.

Next, we evaluate $P_{eq}$ as follows:
\begin{align}
    P_{eq}&=\mathbb{P}(R=T, \bm{\mu}=(1,1,1\ldots,1,\mu_T), \mu_T=K-(T-1)>1)\\
    &=\sum_{r=1}^{T_{\max}}\mathbb{P}(T=r, \mu_1=\mu_2=\ldots=\mu_{r-1}=1, \mu_r=K-(r-1))\\
    &=\left(\sum_{r=1}^{T_{\max}} \sum_{j=1}^{\tau-1}\binom{\tau-j}{r-1}\right)\times \bigg(\frac{1}{\tau}\bigg)^K.
\end{align}
The values of $P_{eq}$ for different values of $\tau$ and $M$ are given in Table~\ref{Peq}. The probability of equal communication costs is negligible and decreases as $\tau$ increases.

\begin{table}[h]
    \centering  
    \begin{tabular}{|c|c|c|c|c|c|}
    \hline
    $P_{eq}(\times 10^{-3})$ &   $\tau=2$ & $\tau=4$ & $\tau=6$ & $\tau=8$ & $\tau=10$\\
    \hline
    $M=1$ & $2.93$ & $1.43\times 10^{-2}$ & $1.04\times 10^{-3}$& $2.37\times 10^{-4}$ & $1.02\times 10^{-4}$ \\
    $M=2$ & $2.93$ & $1.43\times 10^{-2}$ & $1.04\times 10^{-3}$& $2.37\times 10^{-4}$ & $1.02\times 10^{-4}$\\
    $M=3$ & $2.93$ & $1.43\times 10^{-2}$ & $1.04\times 10^{-3}$& $2.37\times 10^{-4}$ & $1.01\times 10^{-4}$\\
    $M=4$ & $2.93$ & $1.43\times 10^{-2}$ & $1.04\times 10^{-3}$& $2.37\times 10^{-4}$ & $9.67\times 10^{-5}$\\
    \hline
    \end{tabular}
    \caption{$P_{eq}$ with $K=10$ for different values of $\tau$ and $M$.}
    \label{Peq}
\end{table}

\section{The Case of $N>2$ Entities: Star Setup}\label{achievable>2_star}
For the star setup, our DOFSP scheme is based on sequentially applying the PSI scheme of  \cite{zhushengMultiPSI} on disjoint subsets of $\mathcal{P}_1$, determined by equi-cost elements. The leader $E_1$ chooses the retrieval indices from $X_1$ as $\mathcal{J}_1$ through $\mathcal{J}_R$ where $f_{i_R}$ is the optimum function value in rounds $1$ through $R\in [L]$. The operating field $\mathbb{F}_q$ should satisfy $q>N-1$, where $q$ is prime. To minimize the communication cost, $q$ is chosen to be the minimum such prime that satisfies the above inequality. 

\subsection{Scheme Description}
We illustrate this scheme with the same optimization setting as in Example \ref{example_dofsp_ring}.

\begin{example}\label{example_dofsp_star}
Let $N_i=2$ for all entities, then the candidates for $l^*$ from \eqref{find_lstar} are $\{1,3,4\}$. Out of $E_1,E_3$ and $E_4$, we assign $E_1$ as the leader. Since $f(A)=1$, $f(C)=f(D)=5$, $f(G)=3$ and $f(J)=4$, we have $\mathcal{J}_1=\{3,4\}$, $\mathcal{J}_2=\{10\}$, $\mathcal{J}_3 =\{7\}$ and $\mathcal{J}_4=\{1\}$ with $\mathbb{F}_q$ such that $q=5$. Then, the queries sent in the first round correspond to $\mathcal{J}_1$ and are
\begin{align}
    Q^{(1)}_{i,1}=\{h_1, h_2\}, \quad Q^{(1)}_{i,2}=\{h_1+e_3, h_2+e_4\}, \quad i\in \{2,3,4\},
\end{align}
where each $h_k\in \mathbb{F}_q^{10}$ $k\in \{1,2\}$. The corresponding answers are
\begin{align}
 A^{(1)}_{i,1}&=\{c(h_1^TX_i+S_{i,1}), \ c(h_2^T X_i+S_{i,2})\}, \\
 A^{(1)}_{i,2}&= \{c((h_1+e_3)^TX_i+S_{i,1}+t_{i,2}(1)), \ c((h_2+e_4)^T X_i+S_{i,2}+t_{i,2}(2))\},
\end{align}
where $t_{2,2}(1)+t_{3,2}(1)+t_{4,2}(1)=t_{2,2}(2)+t_{3,2}(2)+t_{4,2}(2)=2$, where $S_{i,j}$, $j\in [2]$ chosen uniformly at random from $\mathbb{F}_5$ by the respective $E_i$ and $c$ is picked uniformly at random from $\mathbb{F}_5\setminus \{0 \}$. Then, $E_1$ subtracts the answers of the first database from the corresponding answers of the second database to get
\begin{align}
    \{c(X_i(3)+t_{i,1}(k)), \ c(X_i(4)+t_{i,2}(k))\}, \quad  i \in \{2,3,4\}, \ k\in \{1,2\}.
\end{align}
Adding these across $i$, $E_1$ obtains
\begin{align}
    Z_3=&c(X_2(3)+X_3(3)+X_4(3)+2)=3c,\\ 
    Z_4=&c(X_2(4)+X_3(4)+X_4(4)+2)=2c,
\end{align}
where both $2c,3c\in \mathbb{F}_5\setminus \{0\}$, implying that both $C$ and $D$ are absent in $\mathcal{P}$. The next best value is assumed by $J$, i.e., $\mathcal{J}_2 = \{10\}$. This time, $E_1$ sends the queries
\begin{align}
     Q^{(2)}_{i,1}=h_3, \quad Q^{(2)}_{i,2}=h_3+e_{10}, & \quad i\in \{2,3,4\},
\end{align}
and receives the answers
\begin{align}
 A^{(2)}_{i,1}&=c(h_3^TX_i+S_{i,3}), \quad A^{(2)}_{i,2}=c((h_3+e_{10})^TX_i+S_{i,3}+t_{i,2}(3)).
\end{align}
By computing $A^{(2)}_{i,2}-A^{(2)}_{i,1}=c(X_i(10)+t_{i,2}(3))$ for each $i$ and adding them yields
\begin{align}
    Z_{10}=c(X_2(10)+X_3(10)+X_4(10)+2)=3c,
\end{align}
which is an element of $\mathbb{F}_5\setminus \{0\}$, implying that $J$ does not belong to $\mathcal{P}$. The third best value is assumed by $G$, i.e., $\mathcal{J}_3 = \{7\}$. Next $E_1$ sends queries
\begin{align}
    Q^{(3)}_{i,1}=h_4 , \quad Q^{(3)}_{i,2}=h_4+e_{7}, 
\end{align}
and receives answers
\begin{align}
    A^{(3)}_{i,1}=c(h_4^TX_i+S_{i,4}),  \quad A^{(3)}_{i,2}=c((h_4+e_{7})^TX_i+S_{i,4}+t_{i,2}(4)).
\end{align}
for $i\in \{2,3,4\}$ and processes them as before to obtain
\begin{align}
    Z_{7}=c(X_2(7)+X_3(7)+X_4(7)+2)=c(3+2)=5c,
\end{align}
which is $0$ in $\mathbb{F}_5$. This implies that $G$ is an element of $\mathcal{P}$, hence, $\mathcal{P}^*=\{G\}$. There are a total of $8$ queries sent to each of $E_2$, $E_3$, $E_4$, each of which is a $10$-length vector of $\mathbb{F}_q$, resulting in $U_{star}=3\cdot 8\cdot 10 = 240$, and there are $8$ corresponding answers, giving $D_{star}=2(N-1)(\alpha_1+\alpha_2+\alpha_3)=24$, hence $C_{star}=(10+1)D_{star}=264$.

Now, if $N_i=3$ for the non-leader entities, the queries sent $i\in \{2,3,4\}$ are
\begin{align}
    & Q^{(1)}_{i,1}=h_1 , Q^{(1)}_{i,2}=h_1+e_3, Q^{(1)}_{i,3}=h_1+e_4 ,\\
    & Q^{(2)}_{i,1}=h_2, Q^{(2)}_{i,2}=h_2+e_{10},\\
    & Q^{(3)}_{i,3} = h_2 + e_7,
\end{align}
and the corresponding answers are as follows,
\begin{align}
 A^{(1)}_{i,1}&=c(h_1^TX_i+S_{i,1}), \\
 A^{(1)}_{i,2}&= c((h_1+e_3)^TX_i+S_{i,1}+t_{i,2}(1)), \\
 A^{(1)}_{i,3}&= c((h_1+e_4)^TX_i+S_{i,1}+t_{i,3}(1)), \\
 A^{(2)}_{i,1}&=c(h_2^TX_i+S_{i,2}), \\
 A^{(2)}_{i,2}&=c((h_2+e_{10})^TX_i+S_{i,2}+t_{i,2}(2)), \\
 A^{(3)}_{i,3}&=c((h_2+e_{7})^TX_i+S_{i,2}+t_{i,3}(2)),
\end{align}
giving $D_{star}=3\lceil\frac{3\times3}{3-1}\rceil=15$ and $C_{star}=(10+1)\cdot D_{star}=165$, which is lower than the $N_i=2$ case.
\end{example}

We now describe the steps of the general achievable scheme.

\paragraph{Assignment of Common Randomness:} The randomness symbols are assigned for $P_1$ indices to each of the non-leader databases, prior to the start of communication in the same manner as in \cite{zhushengMultiPSI}. 
\begin{itemize}
    \item First, $\lceil\frac{P_1}{N_i-1}\rceil$ \emph{local common randomness} symbols uniformly picked from $\mathbb{F}_q$ are shared locally among the databases of the respective non-leader entities. $E_i$ shares the set $\mathcal{C}_{loc,i}=\{S_{i,1},\ldots, S_{i,m_i}\}$ where $m_i=\lceil\frac{P_1}{N_i-1}\rceil$ $i\in [2:N]$.
    \item Then, each database generates a set of \emph{correlated randomness} symbols from $\mathbb{F}_q$. Specifically, database $j\in [2:N_i]$ of $E_i$ generates the set $\mathcal{C}_{cor,i,j}=\{t_{i,j}(1), t_{i,j}(2), \ldots, t_{i,j}(l_{i,j})\}$, where $l_{i,j}=\big\lceil\frac{P_1-j+1}{N_i-1}\big\rceil$ with $t_{i,1}(k)=0$, such that, every $j\in [2:N_i]$, letting
    \begin{align}\label{eq:define_t_hat}
        \hat{t}_{i,l}=t_{i,j}(k), \quad l=(k-1)(N_i-1)+(j-1), k\in \left\{1,\ldots,\bigg\lceil\frac{P_1-j+1}{N_i-1}\bigg\rceil\right\},
    \end{align}
     the following equality
     \begin{align}\label{eq:define_t_hat_sum}
        \sum_{i=2}^N \hat{t}_{i,l}=q-(N-1)
    \end{align}
    is satisfied, for all $l\in [P_1]$.
    \item Finally, the \emph{global common randomness} symbol $c$ chosen uniformly at random from $\mathbb{F}_q\setminus \{0\}$ is known to all entities except $E_1$.
\end{itemize}  

Thus, for each $i\in [2:N]$, $\mathcal{S}_i$ is comprises $\mathcal{C}_{loc,i}, \mathcal{C}_{corr,i,j}, j\in [2:N_i]$  and $c$.

\paragraph{Query and Answer Generation:}  The query and answer generation steps for each round $r$ corresponds to those of the PSI scheme in \cite{zhushengMultiPSI} with the subsets $\alpha_r, r\in [L]$ till $r=R$. We describe it here briefly as follows. We write, $\mathcal{J}_r=\{u_{r,1}, u_{r,2}, \ldots, u_{r,\alpha_r}\}$ with $u_{r,1}<u_{r,2}<\ldots<u_{r,\alpha_r}$. 

In the first round, $E_1$ generates $k_1=\max_{i\in[2:N]}\lceil\frac{\alpha_1}{N_i-1}\rceil$ vectors $h_1, h_2, \ldots, h_{k_1}$ independently and uniformly at random from $\mathbb{F}_q^K$. The queries sent to database $1$ of $E_i$ are
\begin{align}\label{eq:star_query_1}
    Q^{(1)}_{i,1}=\left\{h_1,h_2, \ldots, h_{v_{1,i}}\right\}, \quad v_{1,i} = \left\lceil\frac{\alpha_1}{N_i-1}\right\rceil, \quad i\in [2:N].
\end{align}
Now, $\alpha_1=(v_{1,i}-1)(N_i-1)+l_{i,1}$ with $l_{i,1}<N_i-1$. The queries sent to databases $j\in [2:N_i]$ of $E_i$ for $i\in [2:N]$ are
\begin{align}\label{queries_star}
    Q_{i,j}^{(1)} = \begin{cases}
    \{h_k+e_{u_{1,m}}, k\in [v_{1,i}], m= (k-1)(N-1)+j-1\},& j\in [2:l_{i,1}+1],\\ 
    \{h_k+e_{u_{1,m}}, k\in [v_{1,i}-1], m= (k-1)(N-1)+j-1\},& j\in [l_{i,1}+1:N_i].
    \end{cases} 
\end{align}
The answers that the databases of $E_i, i\in [2:N]$ respond with are,
\begin{align}\label{answers_star}
   A^{(1)}_{i,j} = \begin{cases}
   \{c(h_k^T X_i+S_{i,k}), k\in [v_{1,i}]\},& j=1,\\
   \{c({(h_k+e_{u_{1,m}})}^T X_i+S_{i,k}+{t}_{i,j}(k)), m = (k-1)(N-1)+j-1\},& j\in [2:N_i],
   \end{cases}
\end{align} 
where the values of $k$, for $j\in [2:N_i]$ are the same as those in \eqref{queries_star}. Using these answers, the leader decodes $X_{\cap}(u_{1,m})$ for each $m\in [\alpha_1]$. Let $t_{1,j}(k)$ map to $\hat{t}_{i,l}, l\in [P_1]$ given by \eqref{eq:define_t_hat}, then $E_1$ computes, for each answer from $E_i$, the set of differences, $\forall j\in [2:N_i]$,
\begin{align}
    A_{1,j}^{(1)}-A_{1,1}^{(1)}&=  \{c({(h_k+e_{u_{1,m}})}^T X_i+S_{i,k}+\hat{t}_{i,l}-c(h_k^T X_i+S_{i,k})\}\\
    &=\{c(X_i(u_{1,m})+\hat{t}_{i,l})\}.
\end{align}
Next, using the answers from all $E_{2:N}$, for each $u_{1,m}$, $E_1$ computes the following sum in $\mathbb{F}_q$,
\begin{align}
    Z_{u_{1,m}}&= \sum_{i=2}^{N} c(X_i(u_{1,m})+\hat{t}_{i,l})\\
    &= c\left(\sum_{i=2}^{N} X_i(u_{1,m})\right) + q-(N-1),
\end{align}
from \eqref{eq:define_t_hat_sum}. Now, if $\sum_{i=2}^{N} X_i(u_{1,m})=N-1$, $Z_{u_{1,m}}=0$, which implies $X_i(u_{1,m})=1$ for each $i\in [2:N]$ and $\mathcal{P}_{alph}(u_{1,m})\in \mathcal{P}$. Otherwise, if $Z_{u_{1,m}}\in \mathbb{F}_q\setminus \{0\}$, then $X_i(u_{1,m})=0$ for some $i\in [2:N]$ and $\mathcal{P}_{alph}(u_{1,m})\notin \mathcal{P}$. If $Z_{u_{1,m}}\neq 0$ for every $m\in [\alpha_1]$, then $X_{\cap}$ has a $0$ at all indices in $\mathcal{J}_1$, and the scheme proceeds to round $2$.

In round $r\geq 2$, the queries and answers are denoted by $Q_{i,j}^{(r)}$ and $A_{i,j}^{(r)}$, respectively. These are used to retrieve $X_{\cap}(\mathcal{J}_{r,m})=\{X_{\cap}(u_{r,1}), \ldots, X_{\cap}(u_{r,\alpha_r})\}$. Similar to the DOFSP scheme for $N=2$, queries are assigned using a new random vector or by reusing the previous one, depending on whether the last query was sent to database $N_i$ or database $j\in [2:N_i-1]$, respectively. The same procedure as $r=1$ is followed for both query and answer generation. 

\paragraph{Proof of Privacy:} The privacy guarantee of DOFSP is a direct consequence of the privacy of PSI scheme given in \cite{zhushengMultiPSI}. In each round, every database receives as query, a $K$-dimensional vector chosen uniformly from $\mathbb{F}_q^K$. Each database performs the same computation to determine its answer. 
\paragraph{Communication Cost:} Clearly, round $r$ is equivalent to conducting the PSI on $\mathcal{P}_{alph}(\mathcal{J}_r)$ and the scheme proceeds till $r=R$ where $Z_{u_{R,m}}=0$ for some $l\in \alpha_R$ for the first time. Let $\mathcal{J}^*=\{u_{R,m}: Z_{u_{R,m}}=0\}$, then the solution set is $\mathcal{P}^*=\mathcal{P}_{alph}(\mathcal{J}^*)$. The scheme is an iterative implementation of the PSI on $\sum_{r=1}^R{\alpha_r}$ elements, and hence, results in the download cost,
\begin{align}
    D_{star}=\sum_{i=2}^N \Big\lceil{\frac{(\sum_{r=1}^R \alpha_r) N_i}{N_i-1}}\Big\rceil.
\end{align}
The total communication cost is, therefore, $C_{star} = (K+1)D_{star}$ symbols of $\mathbb{F}_q$. The entities cannot share prior knowledge of $\bm{\alpha}$ as it would violate $E_1$'s privacy constraint. Hence, the communication cost depends heavily on the choice of $E_1$ and is upper-bounded by $\sum_{i=2}^N\lceil\frac{P_1N_i}{N_i-1}\rceil$. 

\begin{remark}
    Similar to the $N=2$ case, $E_1$ skips the PSI step for $\mathcal{J}_R$ if $R=L$ and $\alpha_L=1$, in which case $D_{star}=\sum_{i=2}^N \Big\lceil{\frac{(\sum_{r=1}^{L-1} \alpha_r) N_i}{N_i-1}}\Big\rceil$.
\end{remark}

\subsection{Comparison with the Naive Scheme} 
Note that the condition under which $\mathcal{I}_{nom,i}$ for entities $E_i, i\in [2:N]$ is equal to that of PSI is when all the elements of $\mathcal{P}_1$ have been checked for their presence in $\mathcal{P}$, hence $E_1$ learns $\mathcal{P}^*=\mathcal{P}$. This occurs whenever $R=L$, i.e., when the optimal value of $f$ is $f_{i_L}$, realized by the $M$ elements in $\mathcal{P}^*$. In terms of communication cost, it suffices to compare the download cost $D_{star}$ with $D_{PSI,star}$ in \cite{zhushengMultiPSI}, which is,
\begin{align}
    D_{PSI,star} = \sum_{i=2}^N \left\lceil\frac{P_1 N_i}{N_i-1}\right\rceil,
\end{align}
under the setting where $E_1$ is the leader. We have $D_{star}<D_{PSI,star}$ and equal only when $\sum_{r=1}^R \alpha_r=P_1$, i.e., when the search for $\mathcal{P}^*$ continues till round $R=L$ and $\alpha_L>1$.
\begin{figure}[t]
    \centering
    \includegraphics[scale=0.7]{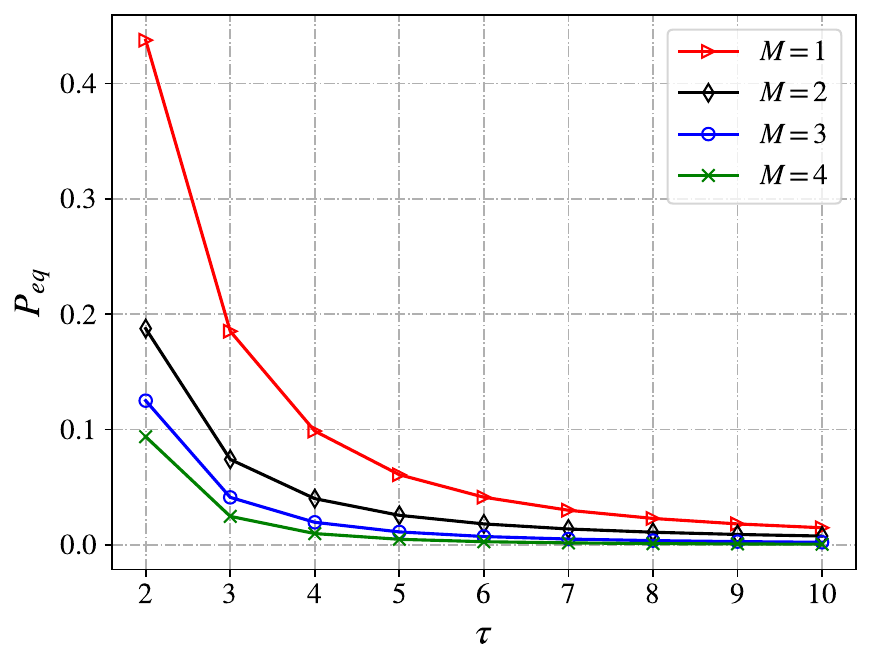}
    \caption{$P_{eq}$ against $\tau$ for various values of $M$ with $N>2$.}
    \label{ProbTMgenN}
\end{figure}
Note that $L$ depends on the realization of $f$ and given the feasible sets $\mathcal{P}_1,\mathcal{P}_2, \ldots, \mathcal{P}_N$, its maximum value is $L_{\max}=\min(\tau,P_1-\max(2,M)+1)$. For $\tau\leq P_1-\max(2,M)+1$, indeed the maximum value of $L$ is $\tau$. For $\tau>P_1-\max(2,M)+1$, this can be verified as follows. If $M=1$, the maximum number of rounds that the communication takes place is $P_1-1$, when $\alpha_r=1$,  $r\in [L]$ throughout which the set intersection is empty. Otherwise, the maximum number of rounds is $L=P_1-M+1$, when $\alpha_r=1$, $r\in [L]$ throughout which the set intersection is empty. 

Under uniform mapping of $f$, $P_{eq}$ is computed as follows,
\begin{align}
    P_{eq}&=\mathbb{P}\left(\sum_{r=1}^{L_{\max}} \alpha_r = P_1, \alpha_L>1\right)\\
    &=\bigg(\frac{1}{\tau}\bigg)^{P_1}\times\sum_{r=1}^{L_{\max}}  \mathcal{N}_{\bm{\alpha}}(r),
\end{align}
where
\begin{align}
   \mathcal{N}_{\bm{\alpha}}(r)= \sum_{j=1}^{\tau-1}\binom{\tau-j}{r-1}\sum_{\alpha_1\geq 1, \ldots,\alpha_{r-1}\geq 1,  \alpha_r\geq \max(M,2):\alpha_1+\alpha_2+\ldots+\alpha_r=P_1}\binom{P_1-M-r}{\alpha_1, \alpha_2, \ldots, \alpha_r-M}.
\end{align}
Given a fixed $P_1$ for the leader set, under varying sizes of $\tau$ and intersection sizes $M$, we plot the probability $P_{eq}$ in Fig.~\ref{ProbTMgenN}. Note the increase in $P_{eq}$ compared to the $N=2$ case given any $\tau$ and $M$. This is because, for the two-user case, $D_{star}$ is always an upper-bound on $D$. 

\section{Conclusion}\label{sec:conclude}
In this work, we proposed an information-theoretic formulation of constrained distributed optimization problem with multiple agents subject to a privacy constraint, i.e., distributed optimization with feasible set privacy (DOFSP). Our DOFSP formulation is closely connected to several private multi-set operations, such as, private set intersection (PSI), PSI cardinality (CarPSI) and threshold PSI (ThPSI), which are of independent interest. The main idea of our formulation is to assign a single leader agent to coordinate the communications, with non-colluding databases of the non-leader agents. At the same time, these databases, with shared randomness, allow zero leakage of information on the feasible sets that they store. As our main result, we show that, compared to a naive scheme where the entities invoke a PSI protocol to learn the entire set intersection, and then compute the solution, our DOFSP scheme is more efficient. Our scheme saves the communication cost with a high probability, and most importantly, prevents additional leakage about the set intersection beyond the solution set. 

For $N=2$, and the star setup, we adopt an SPIR framework for our DOFSP schemes. Unlike the $N=2$ case which evaluates the set cardinalities before finding the intersection, our scheme sequentially finds set intersections for the star setup. This increases the communication cost for $N>2$ since the leader does not know apriori whether the intersection is empty. While the assignment of the leader minimizes the worst-case communication cost for the DOFSP problem, it is not optimized specifically for the latter. On the contrary, the communication cost in the ring setup is independent of the choice of leader. Thus, developing a more efficient DOFSP scheme for the star setup, is open for further investigation. 

For $N>2$ under the ring setup, we utilize secret-sharing and one-time padding to develop our primitives. We observe that the PSI scheme here operates differently from that in \cite{zhushengMultiPSI} adopted in the star setup. The communication cost in the ring setup scales linearly with $N$ and $K$ and is independent of $P_1$, the leader set size, while that in the star setup grows linearly with $P_1$, along with $N$ and $K$. However, the communication cost of the scheme in \cite{zhushengMultiPSI} reduces with more databases per entity, while that in the ring setup is unaffected by this quantity.

In our formulation, we did not make any assumption on the objective function $f$. However, restricting to specific classes of functions can induce special structure on $\bm{\mu}$. This can possibly improve the communication cost by the design of schemes tailored to this structure. Furthermore, it would be interesting to consider more general communication setups, apart from star and ring, considered in this paper. Finally, extension to provide resilience against other threat models, through information-theoretic privacy is another promising direction. In our work, we assumed honest agents, who do not collude among themselves. It would be interesting to study the DOFSP problem with colluding agents that share information and try to breach the privacy of the remaining agents. 

\bibliographystyle{unsrt}
\bibliography{reference}
\end{document}